\documentclass[journal,comsoc,onecolumn]{IEEEtran}
\IEEEoverridecommandlockouts
\usepackage{cite}
\usepackage{amsmath,amssymb,amsfonts}
\usepackage{algorithmic}
\usepackage{graphicx}
\usepackage{textcomp}
\usepackage{xcolor}
\usepackage{multirow}
\usepackage{comment}
\usepackage[hyphens]{url}
\usepackage[hidelinks]{hyperref}
\hypersetup{breaklinks=true}
\usepackage[justification=centering]{caption}
\usepackage{caption}
\usepackage{subcaption}
\usepackage{stfloats}
\usepackage{makecell}
\usepackage{adjustbox}
\usepackage{fancyhdr}
\def\BibTeX{{\rm B\kern-.05em{\sc i\kern-.025em b}\kern-.08em
    T\kern-.1667em\lower.7ex\hbox{E}\kern-.125emX}}

\fancypagestyle{FirstPage}{
\lfoot{LLNL-JRNL-2004219} 
}
\begin{document}

\title{An Integrated Transportation Network and Power Grid Simulation Approach for Assessing Environmental Impact of Electric Vehicles\\
}

%\thanks{Authors would like to gratefully acknowledge support from ...}

\author{\IEEEauthorblockN{Diana Wallison, Jessica Wert, Farnaz Safdarian,\\ Komal Shetye, Thomas J. Overbye, Jonathan M. Snodgrass\\  }
\IEEEauthorblockA{\textit{Department of Electrical and Computer Engineering} \\
\textit{Texas A\&M University}\\
College Station, TX \\
\{diwalli, jwert, fsafdarian, shetye, overbye, snodgrass\}@tamu.edu}\\
\and 
\IEEEauthorblockN{Yanzhi Xu}\\
\IEEEauthorblockA{\textit{ElectroTempo} \\
Arlington, VA \\
 ann.xu@electrotempo.com}

%}

%\thanks{place any grant acknowledgments here}
}
\maketitle

\begin{abstract}

This study develops an integrated approach that includes EV charging and power generation to assess the complex cross-sector interactions of vehicle electrification and its environmental impact. The charging load from on-road EV operation is developed based on a regional-level transportation simulation and charging behavior simulation, considering different EV penetration levels, congestion levels, and charging strategies.  The emissions from EGUs are estimated from a dispatch study in a power grid simulation using the charging load as a major input.  A case study of Austin, Texas is performed to quantify the environmental impact of EV adoption on both on-road and EGU emission sources at the regional level.  The results demonstrate the range of emission impact under a combination of factors.  

\end{abstract}


\section{Introduction}
\thispagestyle{FirstPage}
The wide adoption of electric vehicles (EVs) will fundamentally change the structure of the energy supply in the transportation sector, with a substantial amount of energy coming from the power generation and contributing to higher emissions at power plants. Numerous studies quantify the electricity generation units (EGUs) emissions due to EVs using aggregated emission rates multiplied by the EV charging load.  However, actual emissions depend on dispatched EGUs, which further depend on the EV charging demand, and the characteristics of the power grid.  Therefore, an integrated approach that includes EV charging and power generation is needed to assess the complex cross-sector interactions of vehicle electrification and its environmental impact.

EVs use electricity from onboard batteries that can be charged from an external source \cite{1}.  The use of electric power generated with higher efficiency compared to the internal combustion engine makes EVs a promising alternative for reducing on-road energy consumption \cite{2}.  Among all types of EVs, battery electric vehicles (BEVs) generate zero emissions during on-road operation and therefore will continue to grow in importance for sustainable development. 

From 2000 to 2018, the BEV market share among all light-duty vehicles (LDVs) grew from 0\% to 1.4\%, with roughly 637,000 BEVs in total sold in the U.S. \cite{3}.  For future scenarios, different studies project different but potentially rapid rates of EV usage.  Based on current regulation, technology development, and economic growth, EVs could take at least 20\% market share by 2050, with cumulative EV sales since 2019 reaching 40 million \cite{4}.  Under fast technology diffusion and optimistic market growth, EVs could reach nearly 100\% of new sales and about 45\% of vehicles in operation by 2050 \cite{5}. In any case, EV will be a critical component in future transportation and electricity systems with potential impacts on the grid.

The BEVs generate almost zero tailpipe emissions, while the upstream emissions associated with power generation units (EGUs) and PEVs charging contributes to a substantial amount of overall emissions from BEVs \cite{6,7,8}.  In terms of greenhouse gases (GHG) emissions, the combination of transportation electrification and power generation with renewable energy can potentially yield substantial GHG reduction benefits \cite{9}.  However, if the carbon intensity of EGUs is high, such as in the Midwest region of the U.S., the CO2 emission rates of EVs can be significantly higher than conventional economy cars and even higher than the average light-duty vehicle fleet \cite{10}.  Regarding air pollutants, a recent study finds if 17\% of LDVs, 8\% of heavy-duty vehicles (HDVs), and some off-road equipment can be electrified within the U.S. by 2030, about 3\% of the total NOx emissions can be reduced \cite{6}.  Another study found that under some charging scenarios, the CO emissions with EVs are higher than the emissions under the no-EV baseline \cite{11}.  The actual environmental benefits of wide EV adoption need to be quantified based on the local vehicle operation and electricity generation context.

\subsection{Literature Review}
Due to the lack of a measurement system and the scarcity of real-world data, most existing studies adopt a simulation approach to investigate the impact of EVs on the grid operation and to demonstrate the environmental impact of EVs \cite{9}.  Some of the recent studies that investigated the environmental benefits of EVs focused on the grid have been summarized in Table \ref{tab:table1}.  Those studies found the emissions from EVs depend on various factors, including vehicle type, driving conditions, charging behavior, power generation fuel mix, and pollutant types.  In terms of the modeling approach, most of the existing simulation tools have two analytical components to quantify the EGU-related emissions from EVs \cite{1}: an electricity demand component to quantify EV charging load \cite{2} and an electricity supply component to quantify the power generation to support the EV charging load and the amount of generated emissions.  Depending on the scope of the analysis and the availability of resources, those studies adopt different techniques for each modeling component, which can be broadly categorized as in Table \ref{tab:table1}.

For electricity demand modeling, one dominant method is the distance-based approach, which adopts the average energy rates per unit traveled distance, sometimes adjusted by other factors such as powertrain, vehicle size, and meteorology \cite{6, 11, 12, 13}.  Due to the simplicity of the model and availability of data, this approach is widely used for state-wide or nation-wide analysis.  However, this method may eliminate the impact of the changing traffic operation, which has been identified as an important factor in emissions in other studies \cite{14, 15}.  Another widely used approach for demand modeling is the activity-based approach, which predicts the charging load as a result of vehicle operating and charging activities.  This method reflects the impact of dynamics in transportation systems on electricity demand.  However, this method is more often used in case studies due to the lower scalability and limited data resources.  A demand modeling approach that balances the system variabilities and the scalability to a large region is needed to reveal the impact of EV activities on electricity demand at a larger scale.

On the electricity supply side, the approaches can be categorized by how they obtain the power generation fuel mix, which has been proved by most studies in Table \ref{tab:table1} to have a great impact on the EGU emissions \cite{6, 7, 10, 11, 13, 14, 15}.  The fuel mix can be modeled using historical data, in combination with average or marginal emission rates, to predict the emission impact of EV charging.  As the historical datasets are often publicly available, this method has been widely used to perform large-scale analysis \cite{6, 7, 10, 12, 13}.  However, as electricity generation in some regions is driven by markets, such as Texas, U.S. \cite{16}, the fuel mix is subject to change by market price and may not follow the historical trend.  The fuel mix can also be predicted using a dispatch model, which simulates power generation as a function of real-world electricity load, availability of EGUs, and the grid characteristics to minimize the system operation cost.  This method can reflect the variability of power generation with time and location, but applications are limited to areas where the grid model is available \cite{6, 14, 17, 18}. 
%Also, many dispatch models are only loosely coupled with the EV charging load and limited scenarios are investigated.  It is highly desirable to integrate the dispatch model with the operation-based demand model to fully reveal the environmental impact of EV operation under various implementation scenarios.
\begin{table*}[]
\centering
\caption{Summary of Previous Studies on Environment Impact of EVs}
\label{tab:table1}
\begin{adjustbox}{width=\textwidth}
\begin{tabular}{ccccccc}
\hline
\textbf{Study} &
  \textbf{Scale} &
  \textbf{EV Model} &
  \textbf{Energy demand analysis} &
  \textbf{Energy supply analysis} &
  \textbf{Metrics} &
  \textbf{Results} \\ \hline
{[}11{]} &
  \begin{tabular}[c]{@{}c@{}}State-level \\ (California)\end{tabular} &
  \begin{tabular}[c]{@{}c@{}}PHEVs with 40-mile \\ all-electric range (AER)\end{tabular} &
  \begin{tabular}[c]{@{}c@{}}Distance-based \\ energy modeling\end{tabular} &
  \begin{tabular}[c]{@{}c@{}}predicted hourly \\ dispatch data combined with \\ average emission rates\end{tabular} &
  \begin{tabular}[c]{@{}c@{}}GHG, \\ organic compound \\ and NOX\end{tabular} &
  \begin{tabular}[c]{@{}c@{}}EGU emission changes compared \\ to no EV vary by pollutant \\ types and charging scenarios\end{tabular} \\ \hline
{[}7{]} &
  International &
  \begin{tabular}[c]{@{}c@{}}The Tesla Roadster, \\ the TH!NK   \\ City, and the REVAi\end{tabular} &
  \begin{tabular}[c]{@{}c@{}}Distance-based \\ energy modeling\end{tabular} &
  \begin{tabular}[c]{@{}c@{}}Average emission \\ rates per unit power generation \\ by country\end{tabular} &
  CO2 &
  \begin{tabular}[c]{@{}c@{}}EVs can reduce life-cycle \\ CO2 emissions over the  \\  existing fleet by more than 90\%\end{tabular} \\ \hline
{[}14{]} &
  \begin{tabular}[c]{@{}c@{}}Multiple \\ states\end{tabular} &
  \begin{tabular}[c]{@{}c@{}}PHEVs with different \\ powertrains and AERs\end{tabular} &
  \begin{tabular}[c]{@{}c@{}}Activity-based \\ energy modeling\end{tabular} &
  \begin{tabular}[c]{@{}c@{}}Dispatch model combined \\ with marginal \\ emission rates\end{tabular} &
  \begin{tabular}[c]{@{}c@{}}Energy use\\  and GHG\end{tabular} &
  \begin{tabular}[c]{@{}c@{}}Life-cycle emission reduction  \\ compared to ICEV vary by \\ generation mix and operating conditions\end{tabular} \\ \hline
{[}17{]} &
  \begin{tabular}[c]{@{}c@{}}State-level \\ (Texas)\end{tabular} &
  \begin{tabular}[c]{@{}c@{}}PHEVs with 9.4kWh \\ battery\end{tabular} &
  \begin{tabular}[c]{@{}c@{}}Activity-based \\ energy modeling\end{tabular} &
  \begin{tabular}[c]{@{}c@{}}Dispatch model combined \\ with marginal \\ emission rates\end{tabular} &
  \begin{tabular}[c]{@{}c@{}}CO2,  \\  NOX, SO2\end{tabular} &
  \begin{tabular}[c]{@{}c@{}}Net emission reduction \\ compared to ICEV depend on \\ charging strategies and pollutant types\end{tabular} \\ \hline
{[}15{]} &
  Case study &
  \begin{tabular}[c]{@{}c@{}}4 typical ICEVs and \\ 4 typical \\ 100-mile BEVs\end{tabular} &
  \begin{tabular}[c]{@{}c@{}}Activity-based \\ energy modeling\end{tabular} &
  \begin{tabular}[c]{@{}c@{}}Regression model \\ to predict CO2 emissions \\ under certain grid load\end{tabular} &
  CO2 &
  \begin{tabular}[c]{@{}c@{}}Life-cycle emission reduction \\ compare to ICEV depend on driving cycle, \\ use pattern and generation mix\end{tabular} \\ \hline
{[}10{]} &
  National-level &
  \begin{tabular}[c]{@{}c@{}}One PHEV model \\ (Chevrolet Volt)\end{tabular} &
  \begin{tabular}[c]{@{}c@{}}Assumed charging load\\  from a single EV\end{tabular} &
  \begin{tabular}[c]{@{}c@{}}Marginal emission rates \\ from historic \\ power generation\end{tabular} &
  CO2 &
  \begin{tabular}[c]{@{}c@{}}Emission reduction compared to   \\ ICEV and HEV depend on \\ region and time of charging\end{tabular} \\ \hline
{[}12{]} &
  National-level &
  100-mile BEV (Nissan Leaf) &
  \begin{tabular}[c]{@{}c@{}}Distance-based \\ energy modeling\end{tabular} &
  \begin{tabular}[c]{@{}c@{}}Marginal emission rates \\ from historic \\ power generation\end{tabular} &
  \begin{tabular}[c]{@{}c@{}}Energy use \\ and CO2\end{tabular} &
  \begin{tabular}[c]{@{}c@{}}Temperature and regional fuel mix   \\ have a significant impact \\ on EV CO2 emission rates\end{tabular} \\ \hline
{[}13{]} &
  National-level &
  \begin{tabular}[c]{@{}c@{}}Nissan Leaf, Chevrolet Volt, \\ and Tesla Model S\end{tabular} &
  \begin{tabular}[c]{@{}c@{}}Distance-based \\ energy modeling\end{tabular} &
  \begin{tabular}[c]{@{}c@{}}Marginal emission rates \\ from historic \\ power generation\end{tabular} &
  CO2 &
  \begin{tabular}[c]{@{}c@{}}In most cases, EVs have lower   \\ life-cycle emissions compared to \\ conventional vehicles\end{tabular} \\ \hline
{[}6{]} &
  National-level &
  \begin{tabular}[c]{@{}c@{}}A mix of PHEVs \\ with different ranges \\ and BEV with a \\ 100-mile range\end{tabular} &
  \begin{tabular}[c]{@{}c@{}}Distance-based \\ energy modeling\end{tabular} &
  \begin{tabular}[c]{@{}c@{}}Dispatch model combined \\ with marginal \\ emission rates\end{tabular} &
  NOX &
  \begin{tabular}[c]{@{}c@{}}Total NOX emissions decrease by \\ 209 thousand tons (3\%) \\ under the electrification case in 2030\end{tabular} \\ \hline
{[}18{]} &
  Case study &
  Electric bus &
  \begin{tabular}[c]{@{}c@{}}Activity-based \\ energy modeling\end{tabular} &
  \begin{tabular}[c]{@{}c@{}}predicted hourly dispatch data   \\ combined with \\ marginal emission rates\end{tabular} &
  CO2 equivalent &
  \begin{tabular}[c]{@{}c@{}}The CO2 reduction of the electric buses \\ depends on the time of  \\ charging and power generation by the time of day\end{tabular} \\ \hline
\end{tabular}
\end{adjustbox}
\end{table*}

\subsection{Research Goals}

The major objective of this study is to assess the environmental impact of EV utilization at the regional level.  In particular, the proposed methodology will bridge the following research gaps:

\begin{itemize}
    \item Assess the variability of charging demand under various transportation scenarios. The proposed modeling approach can simulate the dynamic demand changes under different operating conditions instead of using typical operating conditions or historical trends.
    \item Generate on-road and EGU emissions as a result of vehicle operation and charging conditions.  In this case, the environmental benefits at the regional level can be properly assessed.
    \item Provide flexibility in modeling EV penetration, adoption, and charging strategies to answer what-if questions in EV applications.
    \item Measure multiple EV performance metrics, including GHG emissions and critical air pollutants.
\end{itemize}

The proposed methodology will be illustrated with a case study in the state of Texas, where the electric grid is operated by the Electric Reliability Council of Texas (ERCOT) and separated from the rest of the U.S. grid \cite{19}.  
%This study is aimed at generating insights for policymakers and practitioners in Texas regarding the potential impact of EVs under various development scenarios.  Also, the methodology proposed in this study can be adopted by other regions for answering EV-related environmental questions and seeking optimal solutions for promoting the environmental benefits of EVs.

\section{Modeling EV Charging Demand}
\begin{figure*}[htbp]
    \centering
    \includegraphics[width=1\columnwidth]{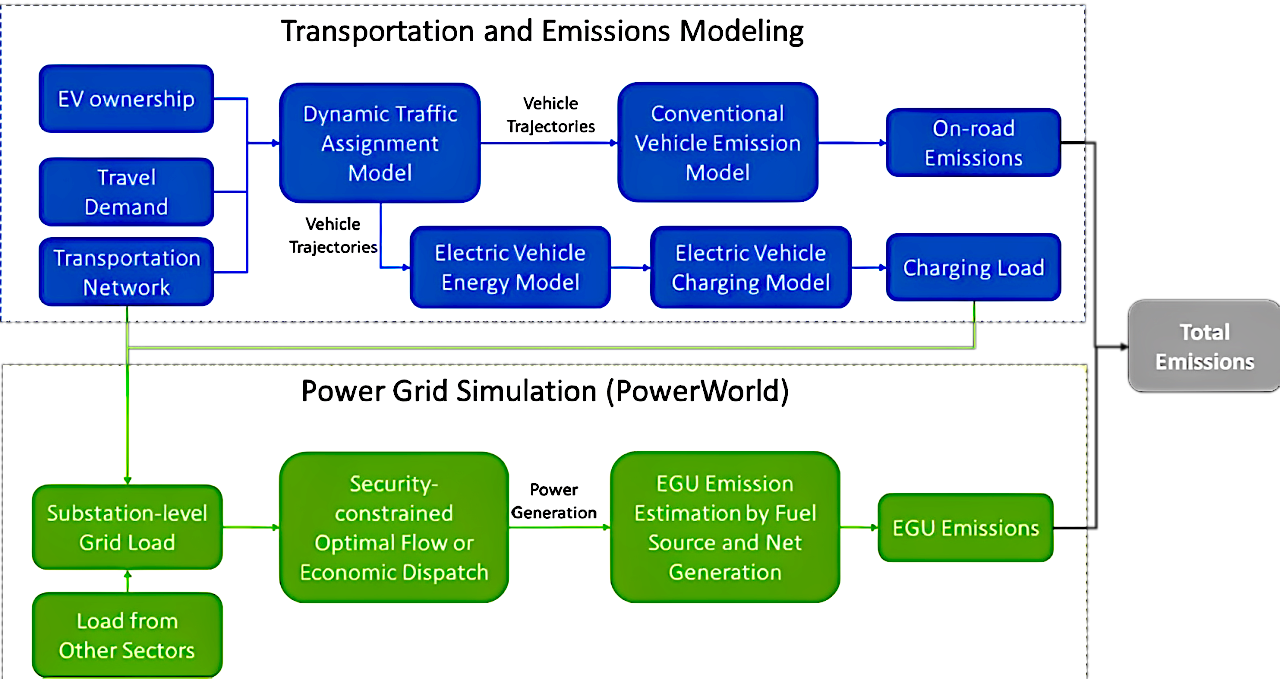}
    \caption{The workflow of the integrated transportation network and power grid simulation}
    \label{fig:workflow}
\end{figure*}

Quantifying the EGU emissions attributed to EV charging is a complex problem and the emission inventory depends on various factors of both transportation systems and grid systems. A methodology that accounts for the system dynamics of both the transportation network and the electric grid is proposed to predict the environmental impact of EV adoption at the regional level.  This proposed methodology is illustrated in Figure \ref{fig:workflow}.  It is composed of two analytical modules – the EV operation and charging analysis module and a power system analysis module, with each module calibrated and validated separately to ensure the level of accuracy in applications.

On the power demand side, a high-fidelity full-chain EV operation and charging model is essential to reasonably predict EV charging loads under various possible system conditions (‘scenarios’) and beneficial for downstream grid analysis.  In this study, the EV energy modeling is performed in parallel with the “Transportation and Emissions Modeling Platform for Optimization” (TEMPO), which is a cloud-based simulation platform that allows rapid integration with other modeling tools (https://tempo-dashboard.io/home) \cite{20}. 

TEMPO facilitates rapid scenario analysis of the system-wide interactions of transportation strategies.  Underlying the platform is a streamlined ensemble of well-recognized models and studies spanning vehicle ownership, travel patterns, energy consumption and emissions,  and pollutant dispersion.  Reduced on-road emissions are estimated using TEMPO by replacing conventional vehicles with EVs. The EV charging load associated with such replacement is estimated based on the EV market share, simulated vehicle trajectories, time-of-day, and other user constraints. The EV charging load is provided as model input for grid simulation.

On the power supply side, the EGU dispatching and fuel supply of power plants under a specific charging load are simulated using the optimal power flow (OPF) method.  
%OPF is a commonly used technique to simulate the generator dispatch decisions in an electricity market environment.  
The OPF method is designed to reflect the major objectives of grid operators, including grid reliability, security, and profitability in real-time. Based on the electricity demand and the conditions of the transmission network, the OPF will calculate the Megawatt (MW) dispatch value of each generator in the system to minimize the operating cost of the system, which includes the cost of power generation \cite{21}.  
In this analysis, the OPF tool available in PowerWorld Simulator \cite{22} is used to simulate the dispatch results in Austin, TX under various EV charging load scenarios. 
%PowerWorld Simulator is an interactive power system simulation package designed to simulate the high voltage power system operation.  It is widely used for power grid planning and simulation \cite{22}. %\color{red} 4.	Why is the formulation of the dispatch model not included in the methods section? (Note: I don't know if the dispatch section was added after this comment or if it was already there and they want equations/diagrams or something like that added, so I kept the comment for discussion) \color{black}

%In the following sections, the major modeling components will be introduced to show the modeling flow in this analysis.  Due to the complexity of each modeling component embedded in this study, only general information about each model is provided here.  Readers are encouraged to use the reference following each tool or concept for technical details and model verifications.

\subsection{Transportation System Modeling}
Grid operators make decisions based on high-resolution information. Therefore, a high spatial and temporal resolution transportation model is needed to supply the spatially resolved electricity demand from EV operation. Mesoscopic dynamic traffic assignment (DTA) models can analyze the movement of individual vehicles while using macroscopic traffic flow theories without complicated vehicle interactions \cite{23}.  

The equilibrium-seeking DTA methods adopt an iterative approach to simulate individual travel behavior under varying traffic conditions and then provide performance measures such as travel time and cost under congestion. The mesoscopic DTA models are highly practical for modeling the dynamics of large-scale networks for long periods at high temporal resolution (e.g., simulate daily travel in a region with a one-minute time interval).  Compared to regional-level travel demand models, which predict average congestion levels for several periods of the day, the DTA offers higher resolution and can capture the delay caused by traffic congestion.  In this case, the charging demand derived from DTA results can be better aligned with power generation from grid simulation.  In this study, a mesoscopic simulation-based DTA model, DynusT \cite{24}, is adopted to perform the DTA analysis. %\color{red} 2.	is the precision afforded by a DTA model necessary to improve the accuracy of energy consumption needed as an input into the grid model?  As opposed to a more simplified travel demand model? 

\color{black}  The DynusT model takes the transportation network and travel demand as inputs and generates vehicle trajectories as output.  The pre-defined EV ownership, represented by the EV market share between 0 and 1, is used to randomly select EV trajectories from all light-duty vehicle trips. %\color{red} Is this a reasonable assumption?  Can trips really be randomly assigned?
\color{black}  The EV assignment at trip-level can be further improved if spatial distributions of EVs are collected through future works.

\subsection{EV Energy Modeling}
In this analysis, the vehicle trajectories generated by DynusT are used to estimate the on-road energy consumption of BEVs.  Using the pre-defined EV market penetration, a fraction of trips is randomly assigned to be EV trips.  As introduced in the literature review, an activity-based vehicle energy model is needed to assess the impact of various transportation-related attributes on energy use.  

In this study, a parameterized simulation-based inference approach is adopted to predict the energy consumption of EVs based on on-road operating conditions \cite{25}.  The vehicle powertrain is represented by a Bayesian Network statistical model, which adopts the domain knowledge as a priori, and can be trained using a data-driven approach.  The details of model development and validation can be found in \cite{25}. 

\begin{figure}[htbp]
    \centering
    \includegraphics[width=0.8\columnwidth]{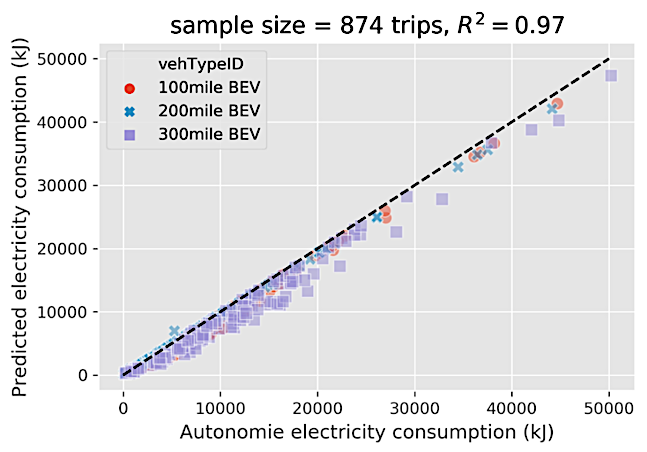}
    \caption{BEV models verification results}
    \label{fig:BEV}
\end{figure}

The study also adopts the energy models for three typical BEVs with 100-mile, 200-mile, and 300-mile ranges, which are most commonly used and analyzed in previous analyses.  The BEV models were validated separately using local driving data collected in the Metropolitan Atlanta area \cite{26, 27, 28} and the full-system vehicle simulation tool called Autonomie. The energy result comparison between the Autonomie and the Bayesian Network model is shown in Figure \ref{fig:BEV}, and the model adopted in this analysis provides prediction accuracy close to the full-system vehicle simulation tool.

Finally, for each assigned EV within the network, the range of the EV is randomly assigned based on the market share of BEVs that are estimated using EV sales data from 2011 to 2019 \cite{29}, assuming EV  ranges are independent of individual trip length. %\color{red} Types of trips and range are likely correlated, but the approach here assumes they are independent.  How can this be addressed in this study? 
\color{black} BEVs with 100-mile, 200-mile, and 300-mile ranges accounted for 25\%, 13\%, and 52\% of the entire BEV fleet.  The energy consumption rates per mile developed from the EV energy models were matched to each trip based on the link-level driving distance and speed.  
%As second-by-second driving profiles are not available in the DynusT output, the driving cycles by different speeds from the EPA MOVES model were used as a surrogate of link driving profiles.  The total on-road energy consumption is then converted to charging demand using the following charging demand simulation.

\subsection{EV charging load modeling}
%By 2019, only 0.12\% of registered vehicles in Texas are BEVs \cite{30}.  Due to the low market penetration and the lack of observed EV charging data, the EV charging demand is estimated under different hypothetical scenarios, with different assumptions made for the spatial and temporal distributions of charging load.  
In this study, the charging load profile hour and location is generated using two methodologies – a simplified method and a realistic method.  The simplified method assumes the charging load equals the trip-level energy use and is charged immediately at the end of the trip.  In this case, the charging load is directly aggregated at the trip end by each hour (referred to as the ‘trip-end’ scenario below).  An off-peak charging profile is also constructed based on the ‘trip-end’ scenario by postponing the charging load assigned to peak hours (2:00 PM – 8:00 PM) to non-peak hours (10:00 PM to 4:00 am of the next day) to reduce electricity cost and peak demand (referred to the ‘off-peak’ scenario below)

The realistic charging demand generation method will simulate the charging demand using a microscopic charging behavior model, which accounts for diversity in people’s range anxiety and the characteristics of daily travel \cite{31}.  The individual-level charging load at different locations depends on the time-of-day, trip characteristics, remaining battery range, and the minimum range needed by individuals to complete trips.  The overnight home charging is assumed for all of the unsatisfied EV charging demand during the day to bring the battery SOC back to 100\% for the next day’s operation. %\color{red} Is it still the case that the vehicle must charge at the end of the day (as in vehicles always start at 100\% SOC)?  Studies have shown that longer-range vehicles may have many days between charging events. 
\color{black} The peak charging load appears around midnight after all vehicles arrive home and start charging. %\color{red} 5.	It was interesting to see the "most likely" charging scenario as peaking after midnight. Particularly interesting in light of the claim (on page 19) that this charging scenario assumes people begin charging as soon as arriving at home. Does this imply that the peak time for arrival at home in Austin is 4 a.m. local time? Please clarify. 
\color{black} In this case, people are more likely to charge by the end of all travels within a day with a battery closer to depletion, instead of charging the vehicle in the middle of the day with sufficient ranges left (referred to as the ‘most likely’ scenario below).  The hourly charging demand will be combined with the electricity load from other sectors (residential, commercial, industrial) to investigate the grid operation and power generation.

\subsection{Grid System Modeling}
A synthetic electric grid that covers the footprint of Travis County, Texas, is used for this study.  The grid model is a combined electric power transmission and distribution (T\&D) network, with the voltage ranging from 120 V (customer level) to 230 kV (high-voltage transmission level).  The power generator information in the model comes from real-world data obtained from the U.S. Energy Information Administration (EIA) \cite{32}, and the location and electricity demand of 307,236 end-use customers are extrapolated from commercially obtained parcel data \cite{33}. The combined synthetic transmission and the distribution network connects the generators and electric loads, and realistically reflects the characteristics of real-world power systems \cite{33}. \color{black} 

The transmission system is validated based on characteristics of actual grids, used as validation metrics \cite{v1,v2},  for achieving realistic data sets; while the distribution system was created and validated using parcel data \cite{33}. The approach uses land parcel data \cite{resstock}, as well as a catalog of technical parameters obtained from commercial and open source data \cite{comstock}. \color{black}

\begin{figure}[htbp]
    \centering
    \includegraphics[width=0.9\columnwidth]{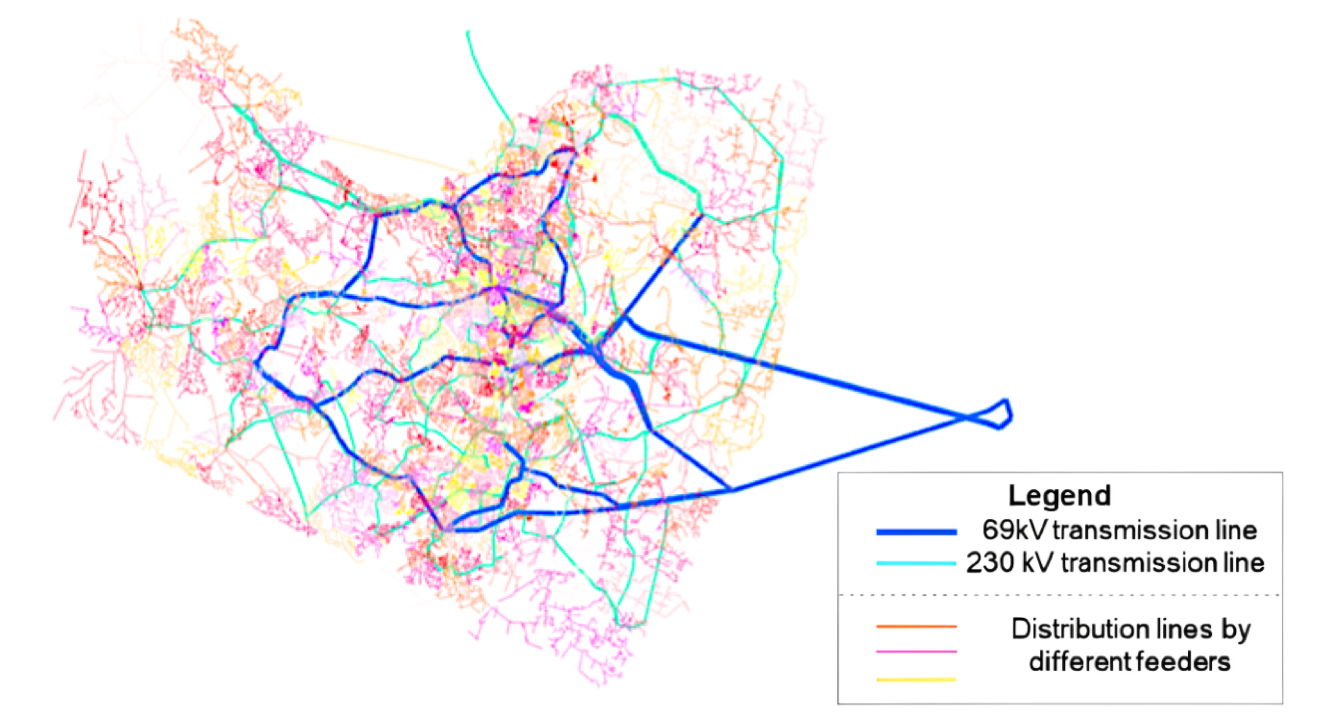}
    \caption{Travis County synthetic grid system \cite{32}}
    \label{fig:Travis}
\end{figure}

For the dispatch analysis, a full T\&D simulation, which can be quite complex and computationally expensive, is not required.  Rather, we use the high-voltage transmission system consisting of 168 nodes (a.k.a. buses) mapped into substations, and the location of those substations are determined to serve major service areas defined by the distribution-level network.  For the distribution system, we use the geographic and topological information from each transmission substation and map it onto the transportation system to estimate the EV load at each transmission substation.  The geographic and electric information of the 39 bulk-level (i.e. transmission level) generation units in this test case are acquired from publicly available data sets.  Figure \ref{fig:Travis} exhibits this Travis County synthetic grid with a total peak load of 3,254 MW. The green and blue lines are the 69 kV and 230 kV transmission networks that constitute the 168-bus test system used in the power grid simulations, with the underlying distribution topology.  Based on the composition of residential, commercial, and industrial customers, hourly load time series is created for each transmission substation in the test case \cite{34} to which the hourly EV load is added, as described in Section 3.1.1 below.  The EGUs were simulated based on the actual power generation information for Austin in 2016 and 2030 and reflect the current and planned practice of a local utility called ‘Austin Energy’ \cite{35}.

\begin{figure}[htbp]
    \centering
    \includegraphics[width=0.9\columnwidth]{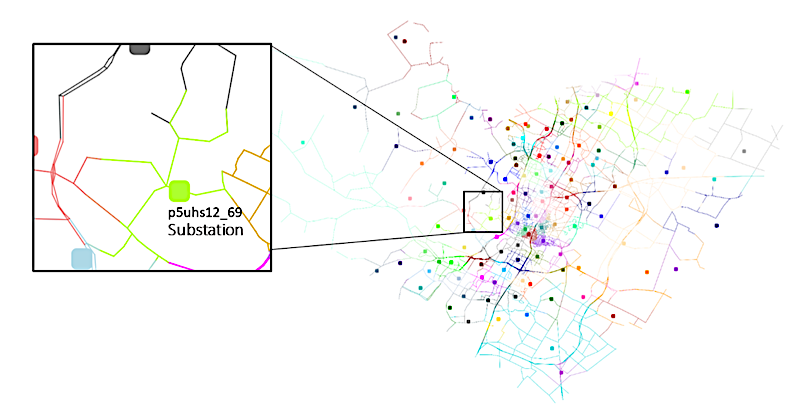}
    \caption{Depiction of transmission substations (dots) and transportation segments (lines) which fall into each substation’s service area, colored by substation service area.}
    \label{fig:serviceareas}
\end{figure}

The transportation network is mapped to the grid by considering the service areas of the transmission substations. These service areas are defined to be the regions spanned by each substation’s distribution-level network. Thus, if a transportation network node lies inside the transmission substation’s service area, it would be serviced by that substation. The mapping process is described in our prevuous work \cite{jessedit2}. Figure \ref{fig:serviceareas} provides a depiction of the transmission substations and the transportation nodes which fall within their service areas.  

For each scenario, the demand from EVs is aggregated to each transmission substation and included in the power grid simulation as a load at that substation. The power generation dispatch is then optimized with the OPF method with combined the EV charging load and the hourly load time series as inputs. The OPF model represents an optimization problem that determines the best operational levels for power plants to meet the electric demands given a transmission network, with the objective of minimizing the operating cost of the system. 

The formulation of an OPF problem is listed below, where N is the total number of nodes in the transmission network, G is the total number of generators, and T is the number of transmission lines. The total cost to dispatch all the generators in the system is the objective function in this formulation. Equations \cite{2,3} are the active and reactive power balance equations, where on each node of the system, the supply and demand of electricity must match. Equations \cite{4,5} reflect inequality constraints from generators, where the output of generators has to be controlled within their designed ranges. Equations \cite{6, 7}  impose constraints on power flow solutions, where the node voltage and transmission line loadings are constrained to be within reasonable ranges for reliable system operation. The output power generation profiles by EGUs are used for the emission calculation as described in the following sections.

\begin{equation} \label{eq:op1}
    min \sum\limits C_i(P_i^G) 
\end{equation}

\begin{equation} \label{eq:Op2}
    s.t.\ P_i=P_i^G-P_i^L \ \forall i \epsilon N
\end{equation}

\begin{equation} \label{eq:Op3}
   \ Q_i=Q_i^G-Q_i^L \ \forall i \epsilon N
\end{equation}

\begin{equation} \label{eq:Op4}
\ P_i^{G,min} \le P_i^G \le P_i^{G,max} \ \forall i \epsilon G
\end{equation}

\begin{equation} \label{eq:Op5}
\ Q_i^{G,min} \le Q_i^G \le Q_i^{G,max} \ \forall i \epsilon G
\end{equation}

\begin{equation} \label{eq:Op6}
\ V_i^{min} \le V_i \le V_i^{max} \ \forall i \epsilon N
\end{equation}

\begin{equation} \label{eq:Op7}
\ L_i \le V_i \le L_i^{max} \ \forall i \epsilon T
\end{equation}

\subsection{Emission Analysis}
In this study, the regional level emissions of fuel combustion are made up of two parts (1) the emissions from the operation on the road of conventional vehicles and (2) the elevated emissions from regional power generation associated with the charging of BEV.  Different methodologies were adopted to quantify the two sources of emissions, respectively, due to different inputs used and different sources of emission rates.

\subsection{On-road emission}
In this study, the on-road emissions were estimated within TEMPO using the MOVES-Matrix, a multi-dimensional MOVES2014a emission rate database \cite{36}. Link speed, link type, and vehicle type distribution generated from the DynusT output are used to query the emission rates.  The link-level emissions were estimated by multiplying the link VMT stratified by time, speed, and vehicle type by the corresponding running exhaust emission rates from the MOVES matrix.

\subsection{EGU emission}
%\color{red} Emissions profiles of EGUs should be summarized, just because they are derived from GREET does not imply that readers have that knowledge first-hand.  Providing some sense, for example, of the capacity mix would immediately provide better context to the readers for the emissions outcomes. \color{black}
In this study, the EGU emissions are estimated using the power generation from the grid simulation and the emission rates from the GREET® model developed by Argonne National Laboratory (ANL) \cite{37}.  GREET® is a life-cycle analysis (LCA) tool that implements a systematic approach to examine the impact of transportation fuels and vehicle technologies on energy consumption and emissions. 
%Compared to other data sources and tools, GREET® has the following advantages:
\begin{comment}
  \begin{itemize}
    \item Provides emission rates for all common GHG emissions and air pollutants, for the entire U.S. with projections to 2050.
    \item Adopts the latest monitored emission data from the U.S. EPA and reflects the real-world power generation conditions across the nation.
    \item Provides emission rates by fuel type and power generation unit type, which can be directly linked to the grid simulation outputs.
\end{itemize}
  
\end{comment}

The emission rates per unit power generation by fuel type were generated from GREET® (data provided in Appendix A). Total energy consumption emissions were calculated using the power generation profiles from grid simulation multiplied by GREET® emission rates.
The results of the simulations of the integrated transportation network and power grid include the emissions on-road and EGU of all vehicles at the regional level.  The results can be used to investigate the environmental benefits of EVs at the regional level and identify key factors of system operation.

\section{Co-Simulation}
%\color{red} How much new load are EVs bringing to the power system in these scenarios? Could easily be derived from the values already in the paper, but would be a notable outcome to highlight.
\color{black}

Currently, due to the low EV adoption rates and the scarcity of charging events, there is no empirical data available for evaluating the performance of models, at the time of this study.  Therefore, the model performance is evaluated using a sensitivity analysis design, which investigates the sensitivity of final emission outputs for various input factors.  The objectives of the sensitivity analysis can be summarized as follows:
\begin{itemize}
    \item Verify the developed energy model to ensure it is sensitive to key operation factors.
    \item Understand the variability of regional emissions under various adoption scenarios.
    \item Determine if the EGU emissions offset the emission reduction from on-road operations at the regional level.
    \item Understand the contribution of operation factors to emissions and help prioritize input preparation.
\end{itemize}

In the following sections, the details of the sensitivity analysis study as well as the power generation and emission results will be introduced. Due to the limited time and resources, only four factors were selected for the sensitivity analysis.  More variables can be included when data sources become available.

\subsection{Input Preparation of Sensitivity Analysis}
In the sensitivity analysis, key factors that have been demonstrated in previous studies to have a significant impact on EV’s energy/emissions were selected.  Those factors include the EV market share, the charging strategy, the static electricity load from other sectors, and ambient meteorology.  The combinations of inputs, summarized in Table \ref{tab:table2}, were considered in the proposed modeling framework.  
%The detailed specifications of each attribute are introduced in the following sections.

\begin{table*}[]
\centering
\caption{Summary of sensitivity analysis inputs}
\label{tab:table2}
\begin{tabular}{lllll}
\hline
\textbf{Attributes}                             & \textbf{Alternative 1} & \textbf{Alternative 2} & \textbf{Alternative 3} & \textbf{Alternative 4} \\ \hline
\textit{\textbf{Weather}} &
  Mild weather with low static   electricity load &
  Hot weather with high static   electricity load &
  \multicolumn{1}{l}{} &
  \multicolumn{1}{l}{} \\ \hline
\textit{\textbf{Power   generation mix}}        & 2016 fuel mix          & 2030 fuel mix          & \textbf{}              & \textbf{}              \\ \hline
\textit{\textbf{EV market penetration in LDVs}} & 5\%                    & 10\%                   & 15\%                   & 20\%                   \\ \hline
\textit{\textbf{Charging   strategies}}         & Trip-end               & Off-peak               & Most likely            &                        \\ \hline
\end{tabular}
\end{table*}

\subsubsection{Weather} 
In Texas, the weather has a substantial impact on electricity demand, and the peak electricity load often occurs during the summer season with high temperatures \cite{38}.  The impact of weather is partially reflected by the static electricity load in this analysis, which represents the electricity demand from the residential, commercial, and industrial sectors.  The high and low load profiles were developed and validated using the real-world electricity data \cite{39} and represent the electricity use under high load from a typical hot summer day and low load from a typical mild weather day in Austin without BEV charging.  The hourly load profiles are illustrated in Figure \ref{fig:pga}.  The 2030 static loads are projected using the demand growth factors from ERCOT \cite{38}, and static electricity load increased by 25\% under both mild- and hot-weather cases.

\begin{figure}[!bp]
  \centering
  \subfloat[]{\includegraphics[width=.45\columnwidth]{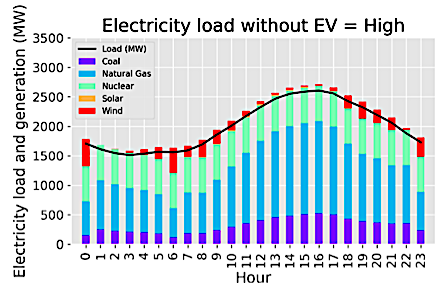}}\label{fig:pga}
  \hfill
  \subfloat[]{\includegraphics[width=.45\columnwidth]{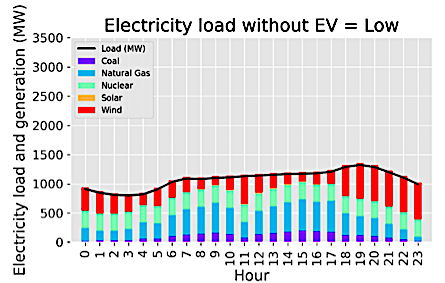}}\label{fig:pgb}
  \hfill
  \subfloat[]{\includegraphics[width=.45\columnwidth]{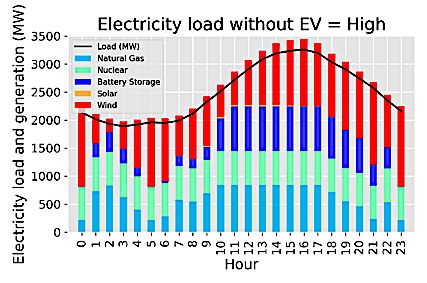}}\label{fig:pgc}
  \hfill
  \subfloat[]{\includegraphics[width=.45\columnwidth]{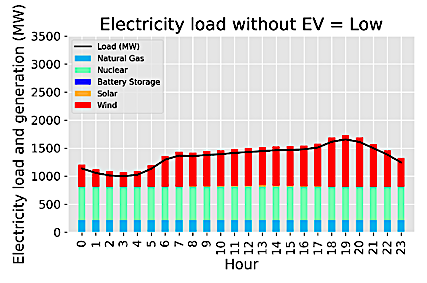}}\label{fig:pgd}
    \hfill
  \caption{Baseline electricity load profiles and dispatch results}
  \label{fig:pga}
\end{figure}

In this analysis, different weather scenarios also reflect different levels of air-conditioning (AC) usage: 1) mild weather case with no air conditioning usage, and 2) hot weather cases with an additional 3 kW air conditioning load for each EV.  Meteorology also affects conventional vehicle emissions. The average July temperature was used for the on-road emissions calculation with AC load, and the average April temperature was used for emissions calculation without AC load. 

\subsubsection{Power Generation Mix}
The power generation mix, represented by the generation capacity by fuel sources, is collected from the long-term plan of Austin Energy \cite{35, 40}.  In the 2016 case, the total power generation capacity is 4376 MW, with around 50\% of power generation from natural gas and 20\% from wind power. 

The 2030 case is updated based on the 2020 test case. %According to Austin Energy’s long-term plan, the Fayette coal plant, and the Decker Creek natural gas plant will retire before 2030.  
3000 MW of wind integration is introduced in the 2030 system to reflect the future power purchase agreement, and 800 MW of battery storage is added to the system.  In the 2030 case, the total power generation capacity is 6541 MW, with 87\% of power generation come from carbon-free resources such as wind and battery storage. The power dispatch results under different power generation mix are also provided in Figure \ref{fig:pga}. The 2030 static loads are projected using the demand growth factors from ERCOT \cite{38}.  

%In the Travis County 2020 case, more than half of the power is generated by natural gas power-plants, and about a third of the power is supplied by carbon-free technologies. The carbon-free energy sources account for greater portions of power generation during the overnight off-peak period. 

In Travis County 2030 case, most of the power is supplied by carbon-free technologies. The power generation from coals will be retired and the share of natural gas decreased significantly under the 2030 case.

\subsubsection{EV market penetration}
In this analysis, we assume four-levels of EV market penetration, ranging from 5\% to 20\% with a 5\% increment.  These EV penetration levels were introduced to determine whether a non-linear relationship exists between the EV market penetration and emissions.

\subsubsection{Charging strategy}
\begin{figure}[!bp]
  \centering
  \subfloat[]{\includegraphics[width=.5\columnwidth]{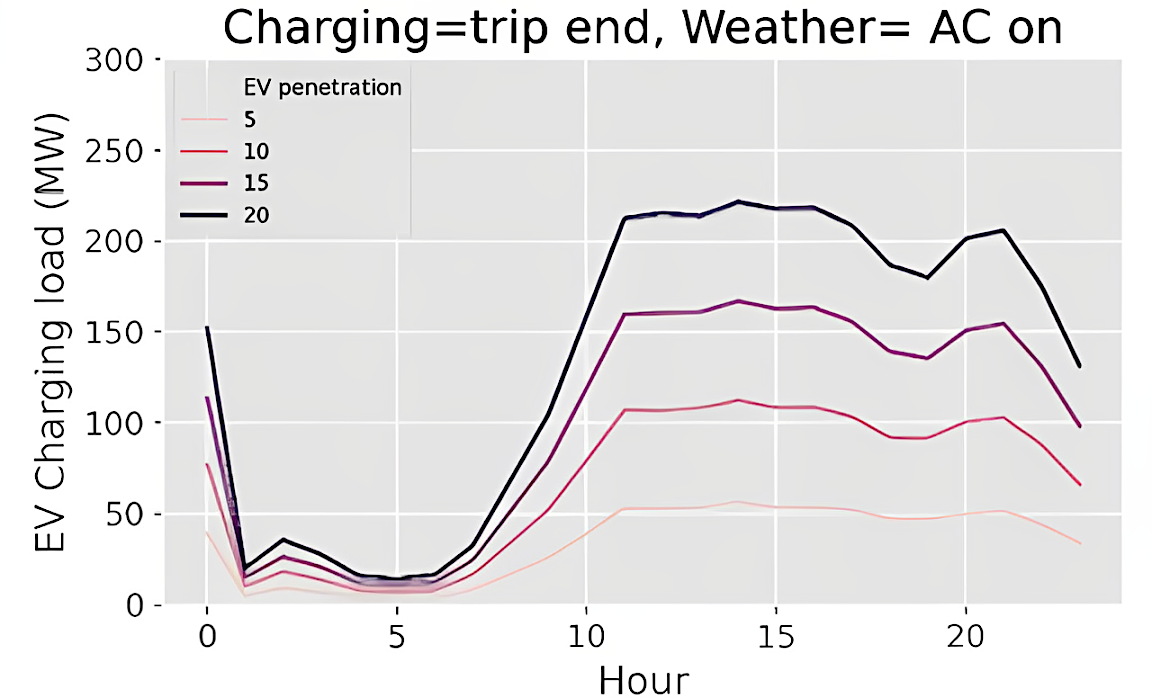}}\label{fig:chargea}
  \hfill
    \subfloat[]{\includegraphics[width=.5\columnwidth]{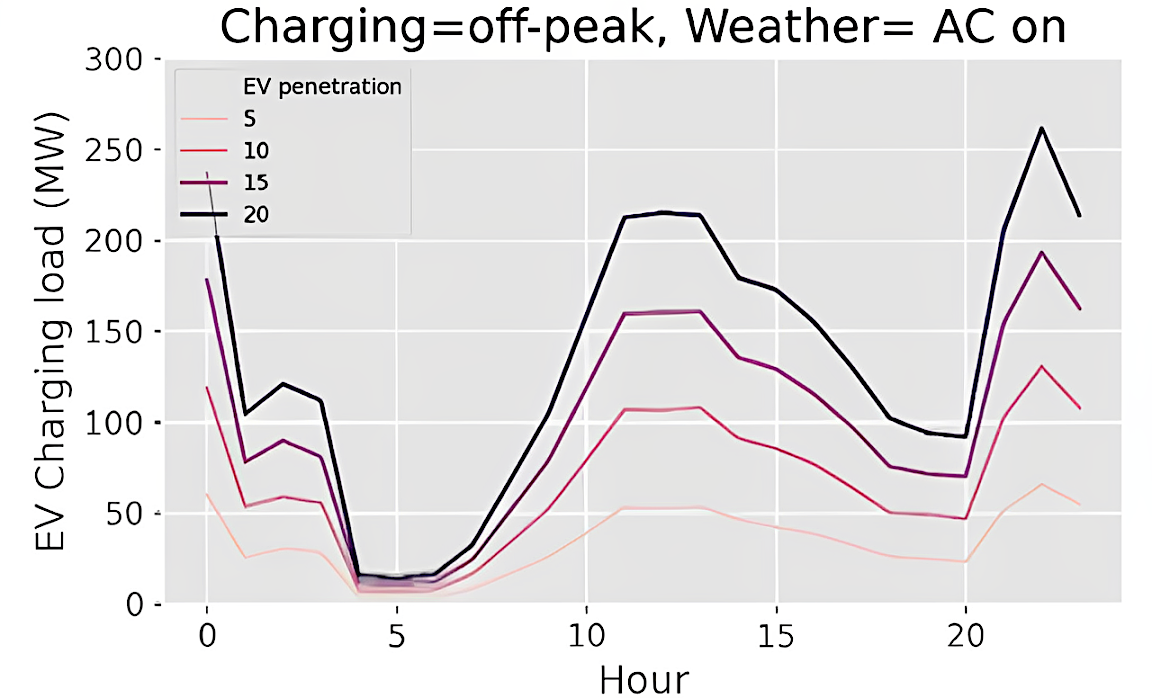}}\label{fig:chargeb}
  \hfill
  \subfloat[]{\includegraphics[width=.5\columnwidth]{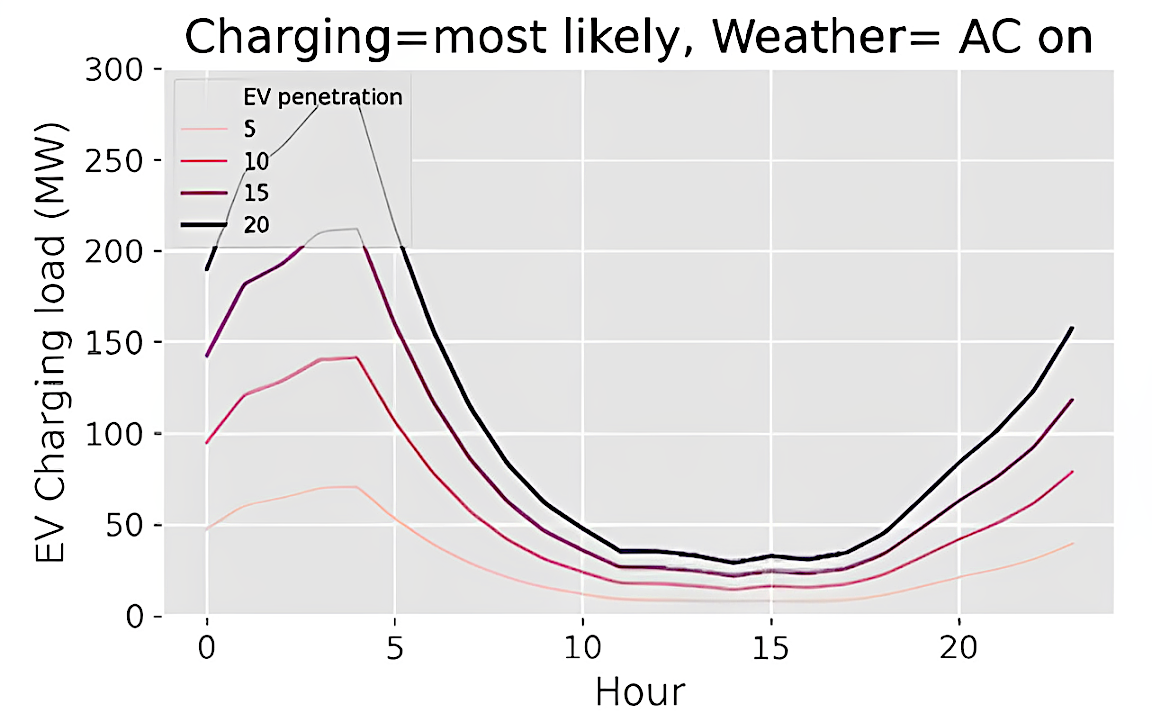}}\label{fig:chargeb}
  \hfill
  \caption{Hypothetical charging profiles under different meteorology}
  \label{fig:chargeprofiles}
\end{figure}

For this study, we use charging profiles used in \cite{jessedit}. The three EV charging profiles generated under different spatial and temporal assumptions on charging behavior introduced previously.  The load due to these charging profiles, combined with different meteorology, is illustrated in Figure \ref{fig:chargeprofiles}.  The ‘most likely’ charging scenario has the highest overnight demand, and the peak load occurred around 3 a.m. with a maximum amount of accumulated overnight charging load. The EV charging demand is higher if the AC load is added.  Using 2020 power generation, the additional EV charging load accounted for 1.5\% - 6\% of all electricity load with 5\%-20\% EV penetration level in the high static load scenario, and 2\%-8\% of all electricity load in the low static load scenario.  Under 2030 power generation, the additional EV charging load accounted for 1 \% - 5\% of all electricity load with 5\%-20\% EV penetration level in the high static load scenario, and 1.5\%-6\% of all electricity load in the low static load scenario.

\begin{comment}
    \begin{figure*}[htbp]
    \centering
    \includegraphics[width=2\columnwidth]{Figures/charging_profiles_high.png}
    \caption{Hypothetical charging profiles under different meteorology}
    \label{fig:chargeprofiles}
\end{figure*}
\end{comment}

In conclusion, the selected input factors represent reasonably diverse EV operation and grid operation scenarios in Texas and, therefore, the resulting emissions should reflect the impact of the considered factors on the total emissions.  In the next section, we summarize the emissions computed for all possible combinations of the inputs and identify the most influential factors.

\subsection{Power Generation by Scenarios}
\begin{figure}[!bp]
  \centering
  \subfloat[]{\includegraphics[width=0.45\columnwidth]{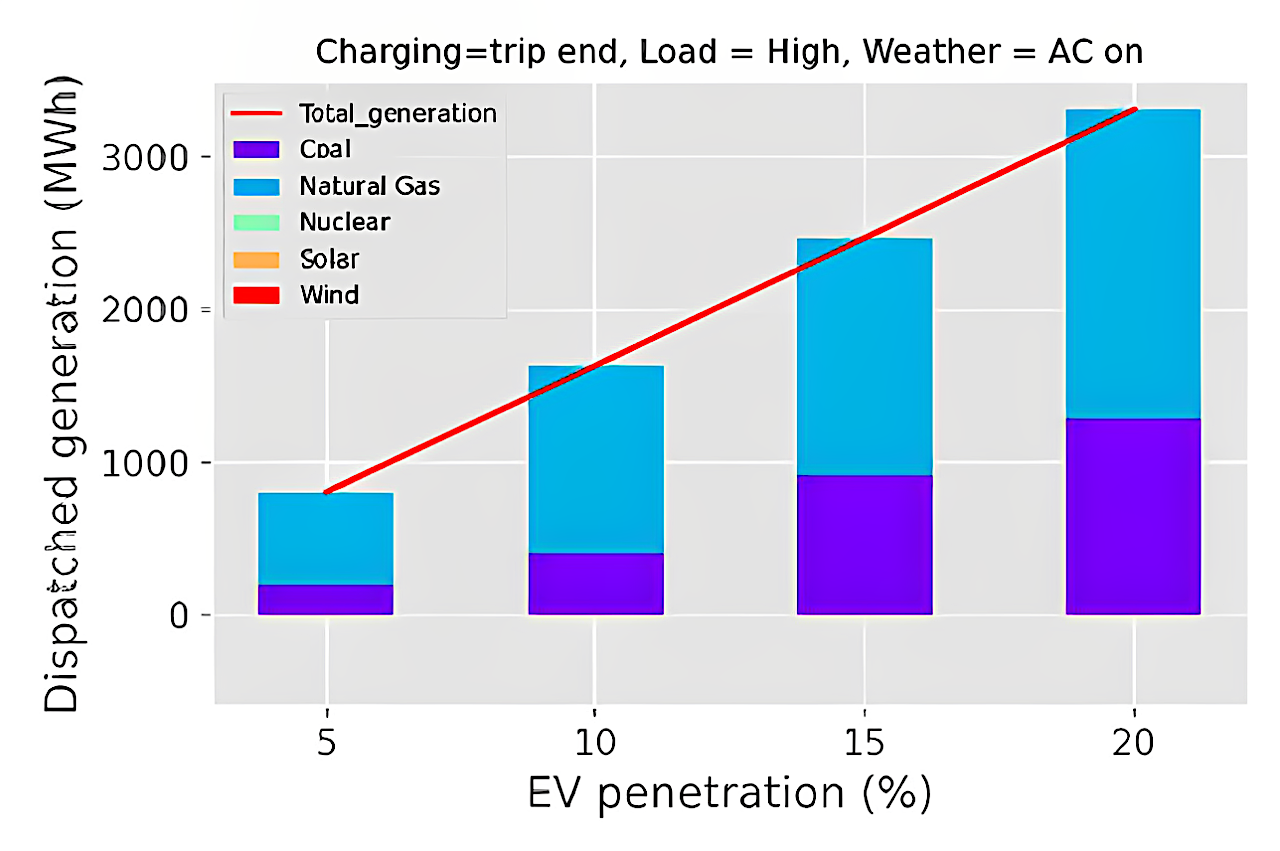}}\label{fig:chargea}
  \hfill
    \subfloat[]{\includegraphics[width=0.45\columnwidth]{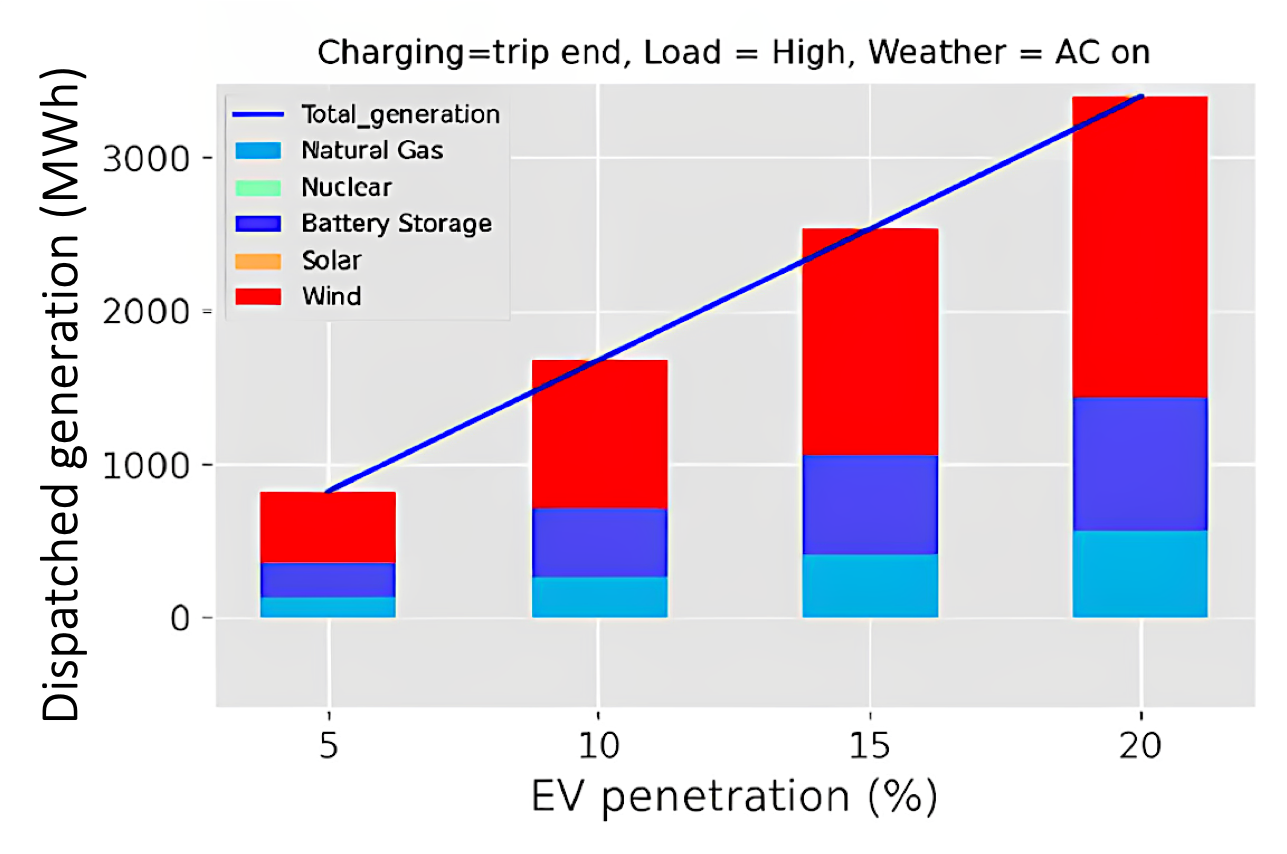}}\label{fig:chargeb}
  \hfill
  \caption{ Trip end marginal power generation under different scenarios: (a)	Power Generation under 2016 Fuel Mix (b) Power Generation under 2030 Fuel Mix}
  \label{fig:charge}
\end{figure}

The power generation profiles under each scenario were compared to understand the impact of EV charging on the power dispatch.  The additional power generation dispatched in response to EV load during a day, represented by the difference in power generation between EV scenarios and no-EV baselines, is shown in Figure \ref{fig:charge}.  

Under the 2016 fuel mix scenario, the EV charging demand is served by natural gas and coal powerplants, with the carbon-free power already dispatched to serve the static load due to its lower costs. The additional power generation increases linearly with the EV market share, but the split between the natural gas and coal power depends on the charging profile and the static electricity load.  

Under the low static electricity load, if a higher EV load is assigned to the overnight period, more natural gas will be dispatched instead of coal due to its low marginal locational costs.  The replacement of coal generation by natural gas is likely to yield environmental benefits, as the natural gas generation emits less GHGs and most types of air pollutants per unit energy generated. 

Under the 2030 fuel mix, most of the additional power generation comes from renewable resources, including wind and battery storage.  When the static load is low under mild weather, most of the additional power generation comes from wind power, which will eliminate emissions from the power sector.  Even under high static load in hot weather, the additional power generation from natural gas accounted for ~10\% of total generation, which suggests much lower emissions from the power sector compare to 2016 cases.

\subsection{EGU Emission Rates by Factors}
\begin{figure*}[!bp]
  \centering
  \subfloat[]{\includegraphics[width=0.45\columnwidth]{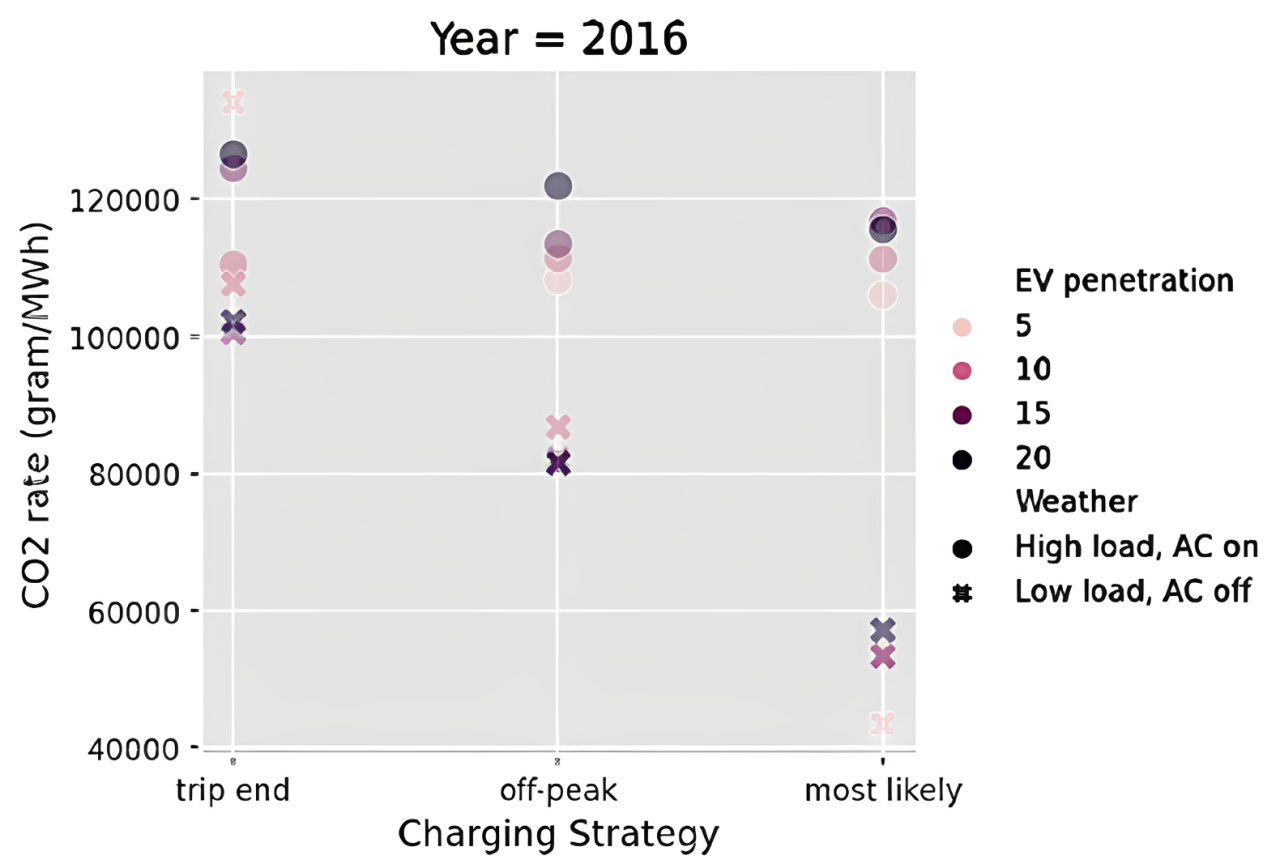}}\label{fig:CO216}
  \hfill
    \subfloat[]{\includegraphics[width=0.45\columnwidth]{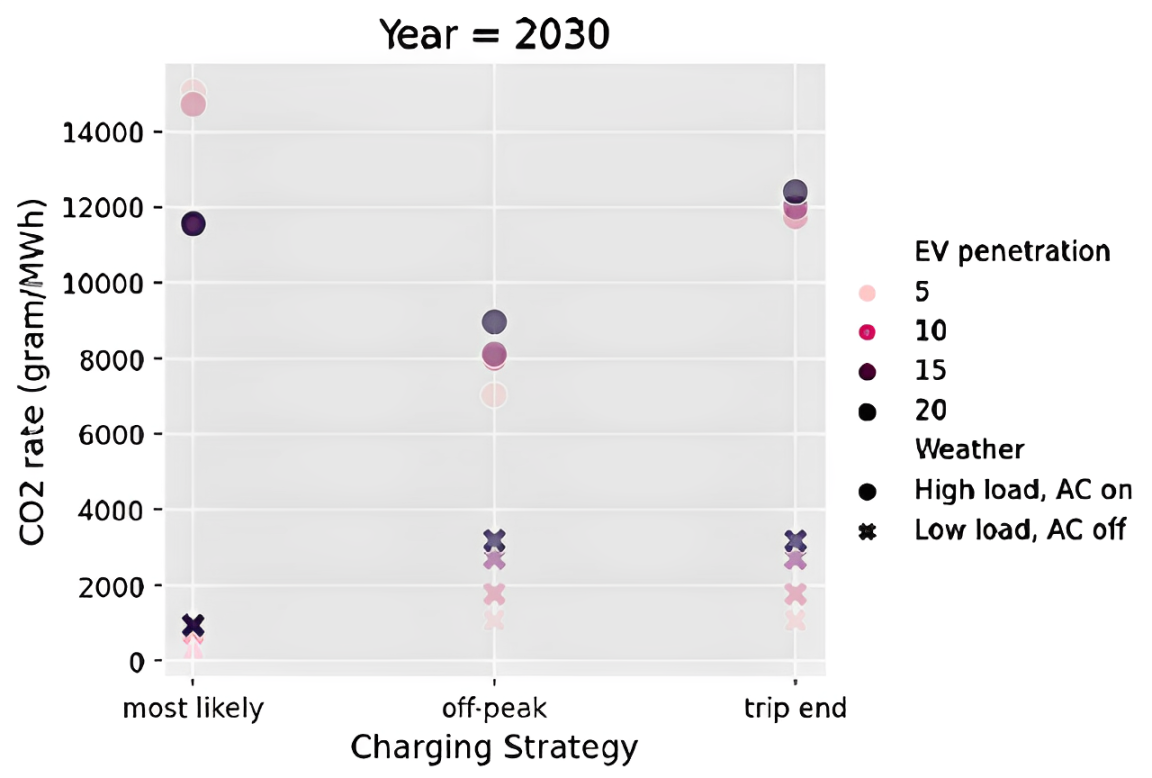}}\label{fig:CO230}
  \hfill
  \caption{Marginal EGU emission rates by factors}
    \label{fig:marginalegu}
\end{figure*}

In this section, the emission rates of marginal dispatched EGUs due to the EV charging load under various scenarios are analyzed to identify the significant factors on the EGU emissions. The scatterplots of the EGU emission rates by each factor and pollutants are shown in Figure \ref{fig:marginalegu}.  Under 2016 power generation, in terms of CO2, NOX, and PM2.5, the emission rates are generally lower under the most likely EV charging load and the low static electricity load because a higher fraction of natural gas is dispatched in those cases. The variation of emission rates is also higher under the most likely EV charging load and the low static electricity load, as a result of the diverging dispatch results under different combinations of those two factors.  

The Volatile Organic Compound (VOC) emission rates are higher under the most likely EV charging load and the low static electricity load, due to higher VOC emission rates of natural gas compared to coal.  The variation of the VOC emission rates is even higher under those two cases due to the higher variation of fuel composition in power generation.  Under 2030 power generation, unlike 2016 results, the emission rates are generally lower under the off-peak EV charging load and the low static electricity load for all pollutants due to a higher fraction of renewable energy dispatched in those cases.  The variation of emission rates is still higher under the most likely EV charging load and the low static electricity load, as a result of the diverging dispatch results under different combinations of those two factors. 

Finally, there is no notable pattern concerning the EV penetration level in all cases, with the emission rates under low and high EV market penetrations falling within similar ranges. In this case, we conclude that the EV charging strategy, fuel mix, and the static electricity load both have a substantial impact on the emission rates, but the direction of the changes depends on the type of emissions investigated.

The spread of marginal emission rates shown in Figure \ref{fig:marginalegu} has revealed intricate interactions among input factors.  For example, our analysis has shown that the most likely charging scenario where charging takes place as soon as people get home is not necessarily an undesirable scenario under the 2016 fuel mix.  With a higher fraction of renewable power generation in 2030, especially wind power during the night, off-peak charging offers greater environmental benefits and align with price signals from utilities.  

The EGU emission rates under the most likely charging strategy display a clear bi-modal pattern for all pollutants, where the emission rates are high under high-load conditions and low under low-load conditions.  This is because, under the low static electricity load and the most likely charging strategy, dirtier power generation (e.g., coal in 2016 and natural gas in 2030)  is replaced by cleaner fuels (e.g., natural gas in 2016 and wind in 2030), which causes the EGU emission rates of CO2, PM2.5, and NOX to decrease.

\begin{comment}
   \begin{figure*}[htbp]
    \centering
    \includegraphics[width=2\columnwidth]{Figures/marginal_egu2.png}
    \caption{Marginal EGU emission rates by factors}
    \label{fig:marginalegu}
\end{figure*} 
\end{comment}

%\color{red} It is unclear, both for vehicle and EGU emissions, if the rates included in this paper can be considered as lifecycle emissions. For consistency, it would be wise to only consider emissions from fuel combustion for both vehicles and EGUs or to consider all lifecycle emissions for both (including upstream emissions associated with fuel extraction, refinement, and delivery). \color{black}

\subsection{Total Emissions by Scenarios}
The total emissions from EGUs and on-road operations were combined to investigate the environmental benefits of EVs.  Although HDVs are assumed not impacted by electrification, the HDV emissions are included as part of regional emission inventories and demonstrate the system-level impacts of LDV electrification.  

\begin{figure*}[!bp]
  \centering
  \subfloat[]{\includegraphics[width=0.45\columnwidth]{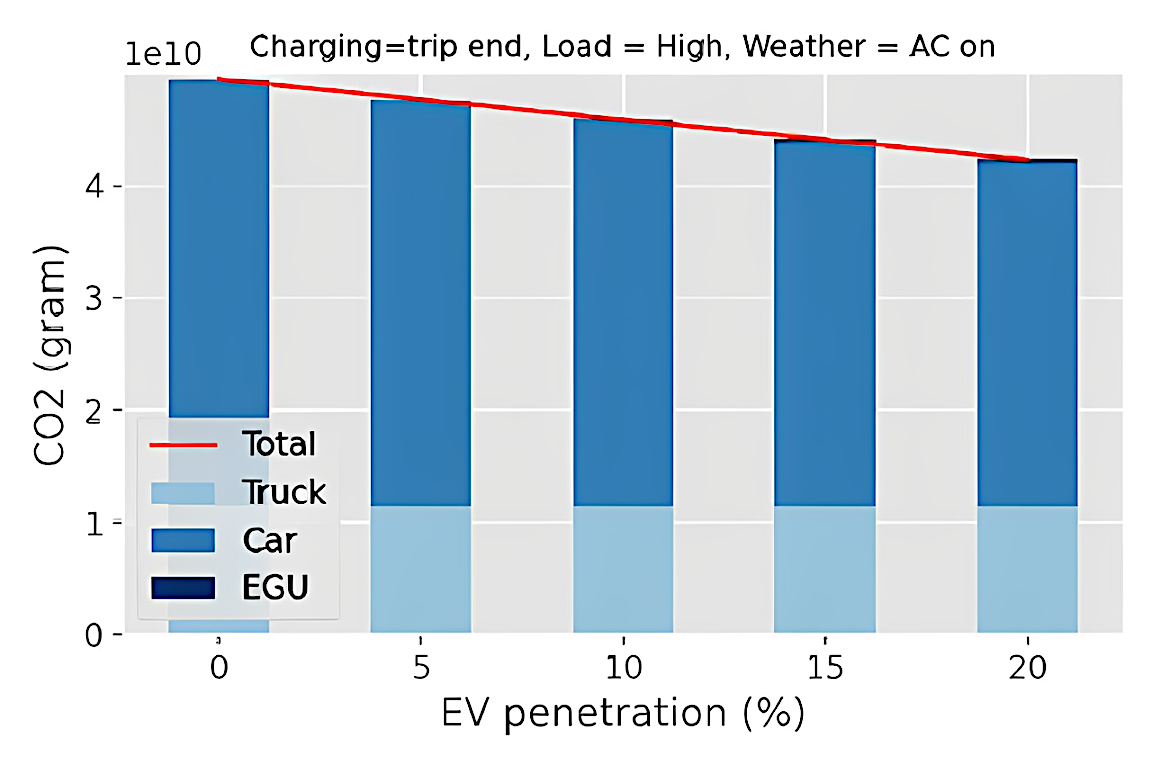}}\label{fig:CO216}
  \hfill
    \subfloat[]{\includegraphics[width=0.45\columnwidth]{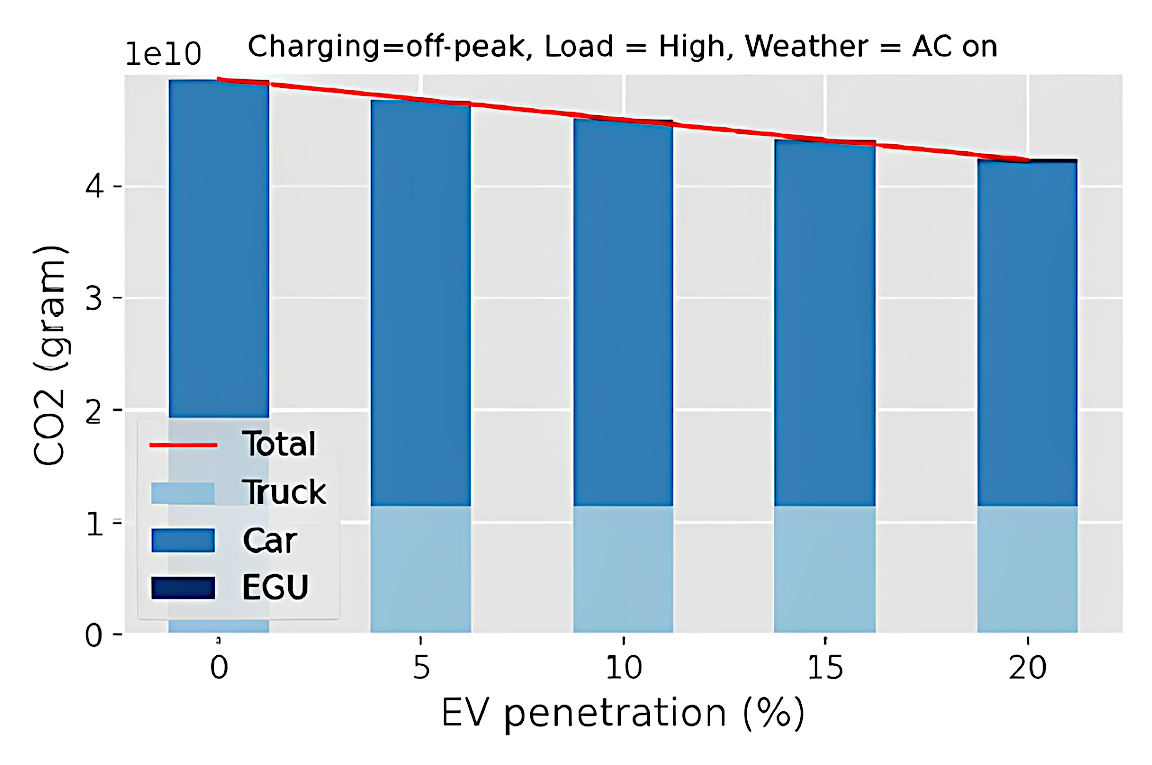}}\label{fig:CO230}
  \hfill
  \subfloat[]{\includegraphics[width=0.45\columnwidth]{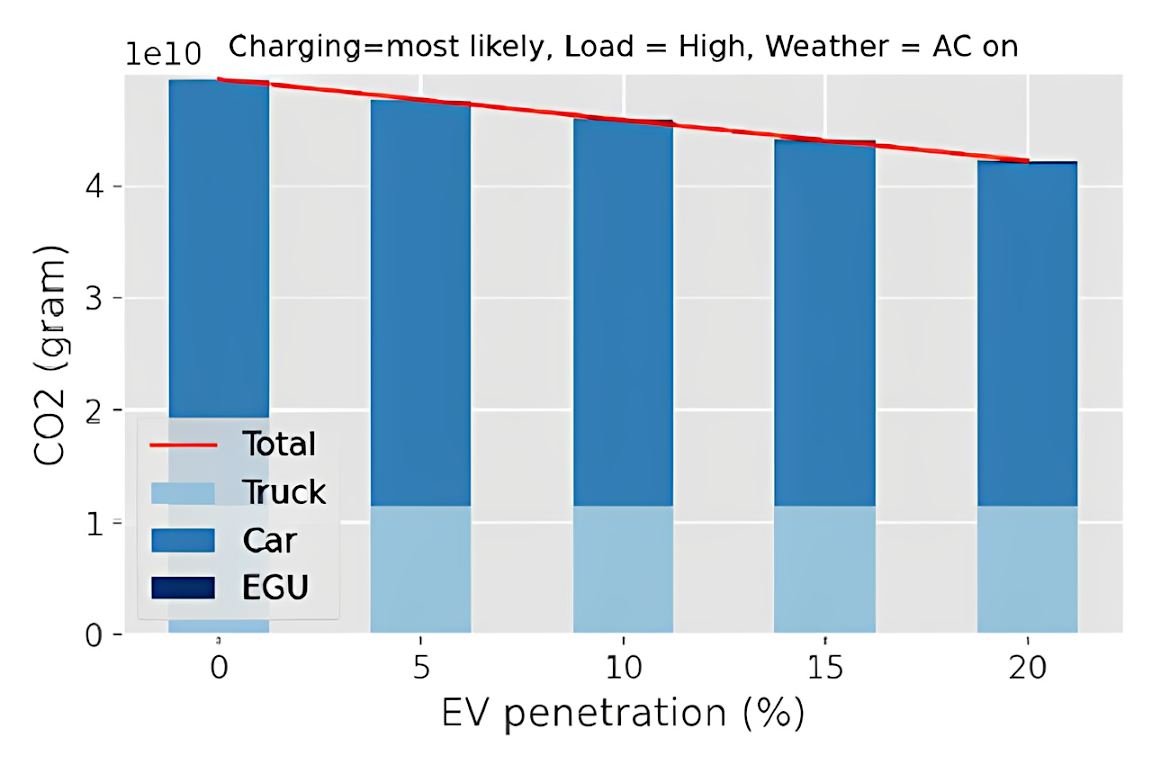}}\label{fig:CO216}
  \hfill
  \subfloat[]{\includegraphics[width=0.45\columnwidth]{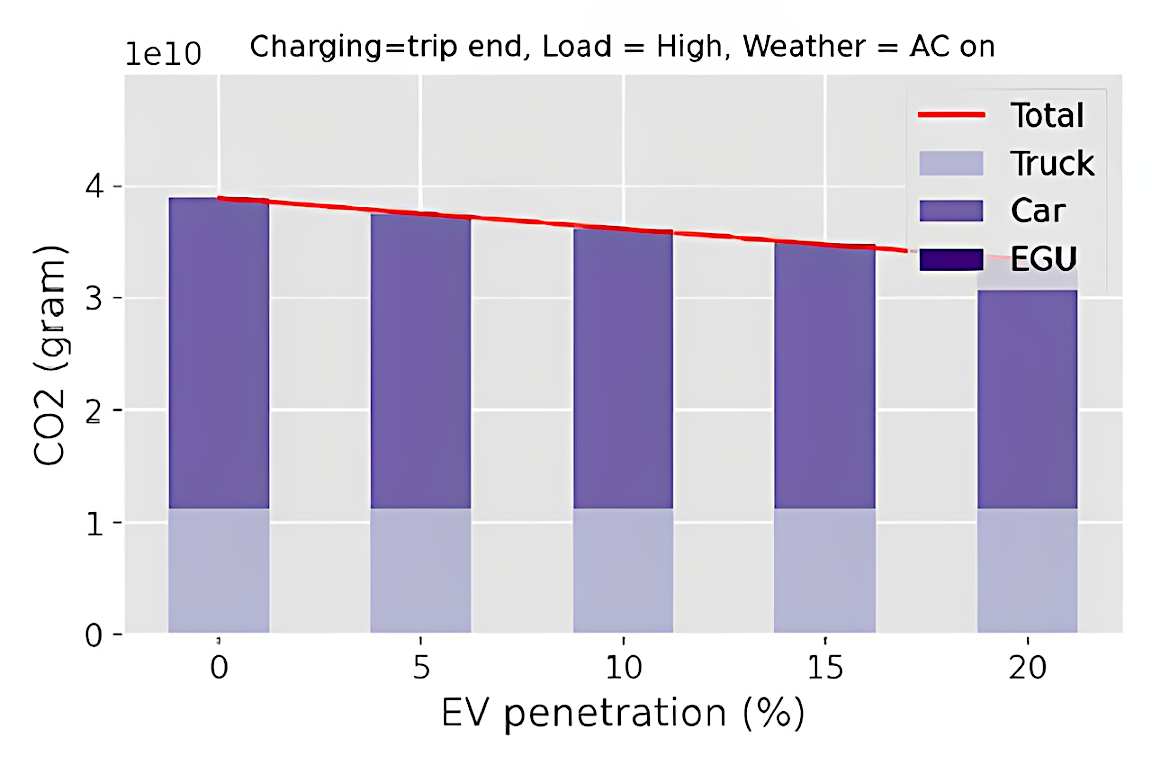}}\label{fig:CO216}
  \hfill
  \subfloat[]{\includegraphics[width=0.45\columnwidth]{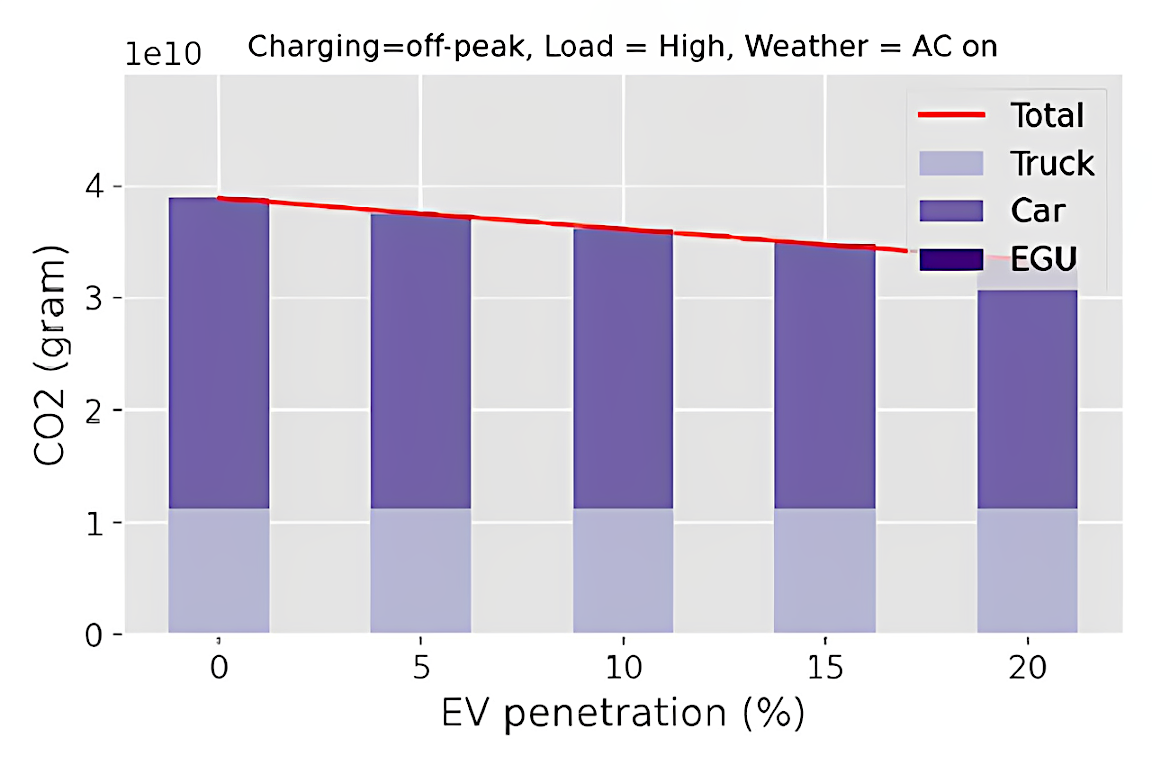}}\label{fig:CO216}
  \hfill
  \subfloat[]{\includegraphics[width=0.45\columnwidth]{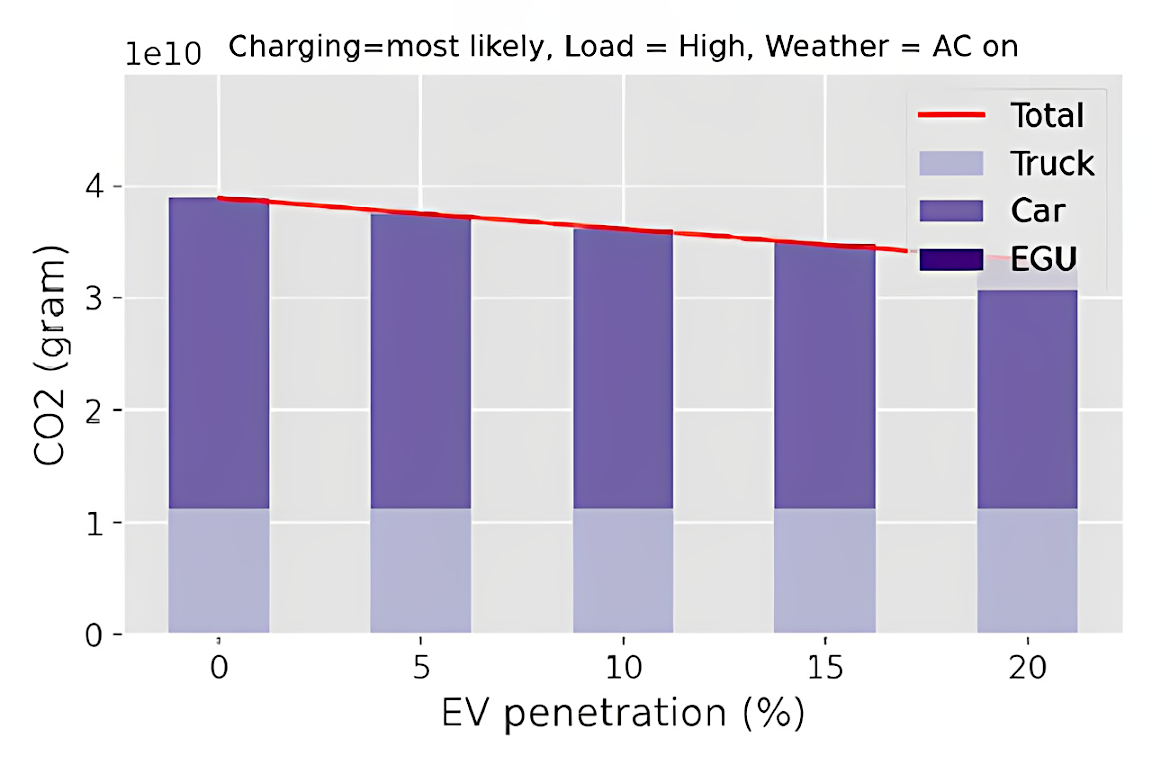}}\label{fig:CO216}
  \hfill
  \caption{Total CO2 emissions by scenario (a,b,c) 2016 Case Total CO2 Emissions by Scenario (d,e,f) 2030 Case Total CO2 Emissions by Scenario}
    \label{fig:CO2}
\end{figure*}

\begin{figure*}[!bp]
  \centering
  \subfloat[]{\includegraphics[width=0.45\columnwidth]{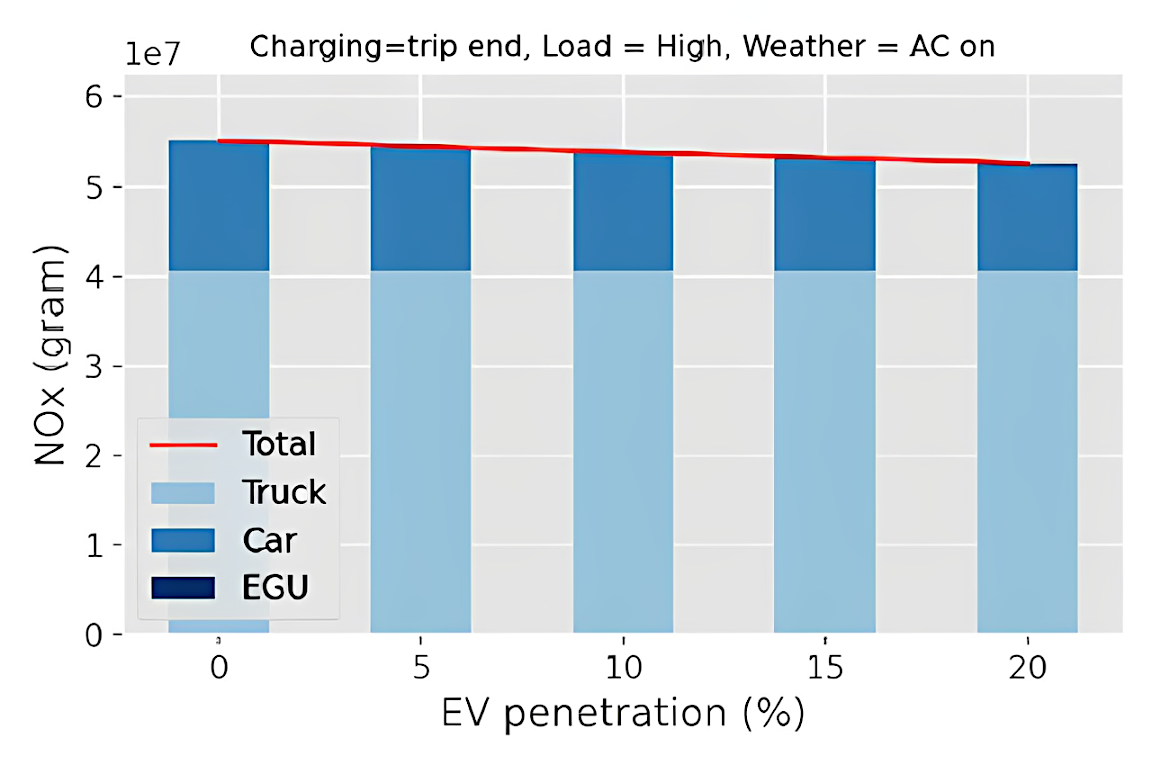}}\label{fig:CO216}
  \hfill
    \subfloat[]{\includegraphics[width=0.45\columnwidth]{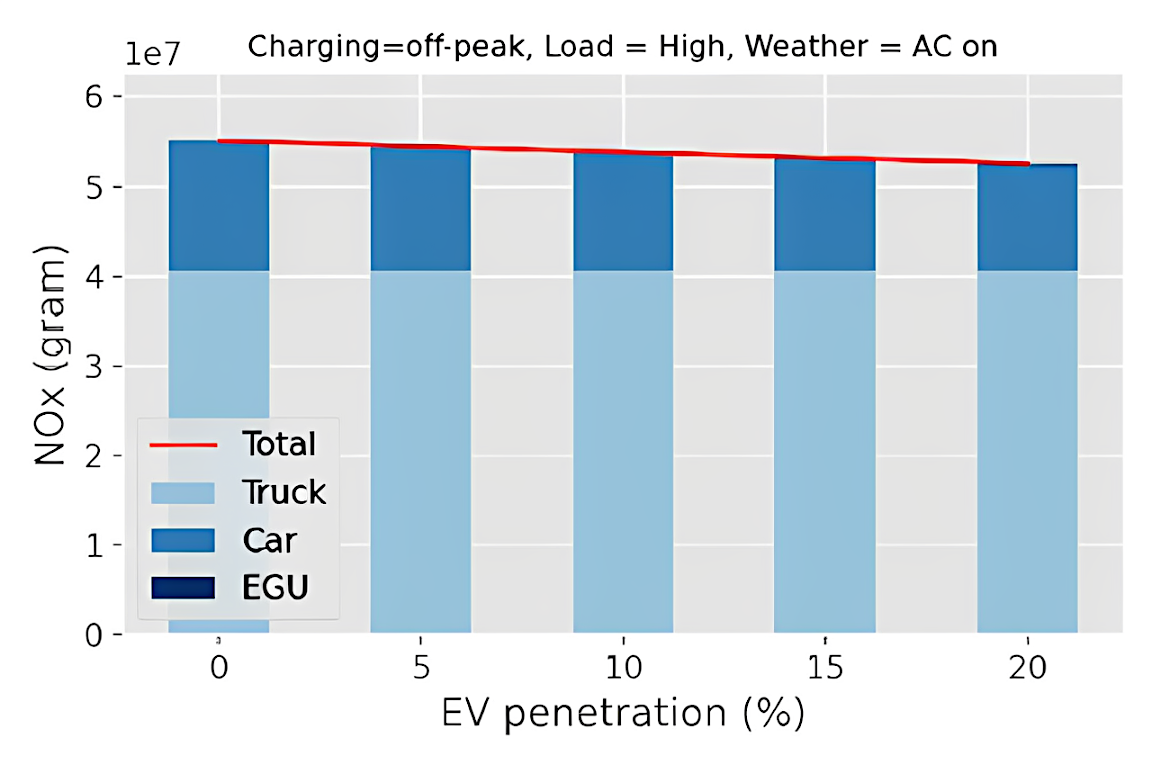}}\label{fig:CO230}
  \hfill
  \subfloat[]{\includegraphics[width=0.45\columnwidth]{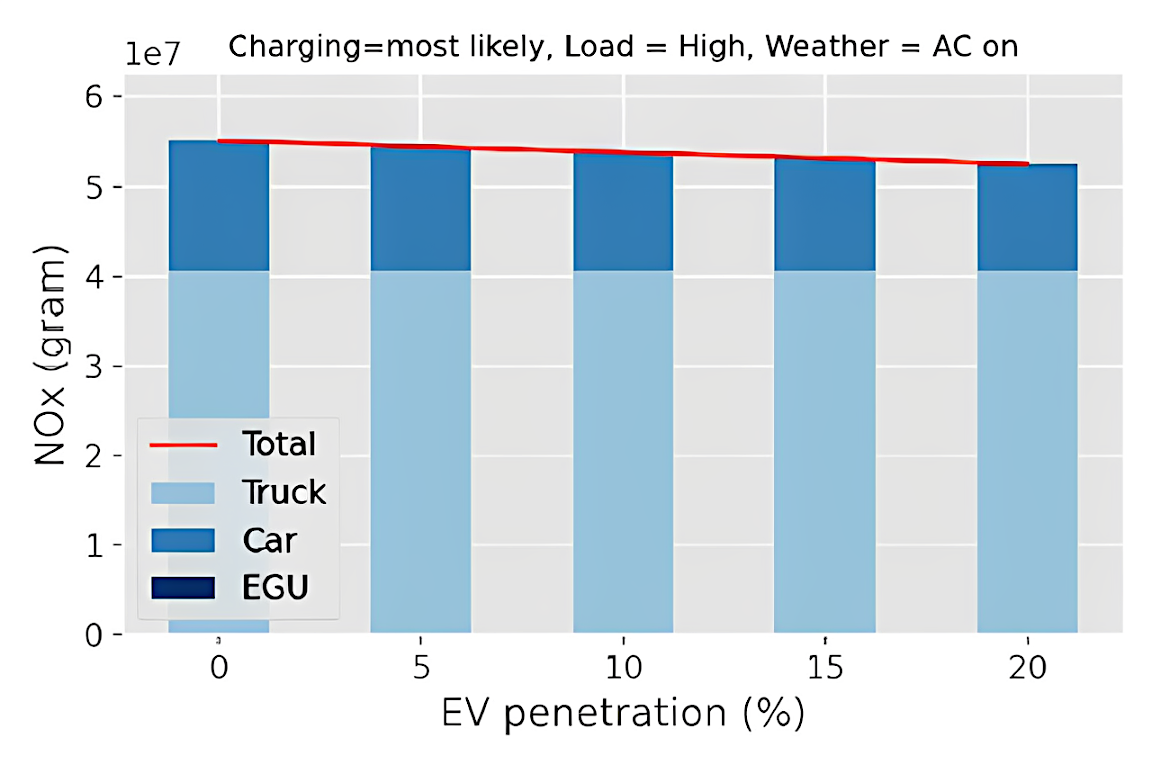}}\label{fig:CO216}
  \hfill
  \subfloat[]{\includegraphics[width=0.45\columnwidth]{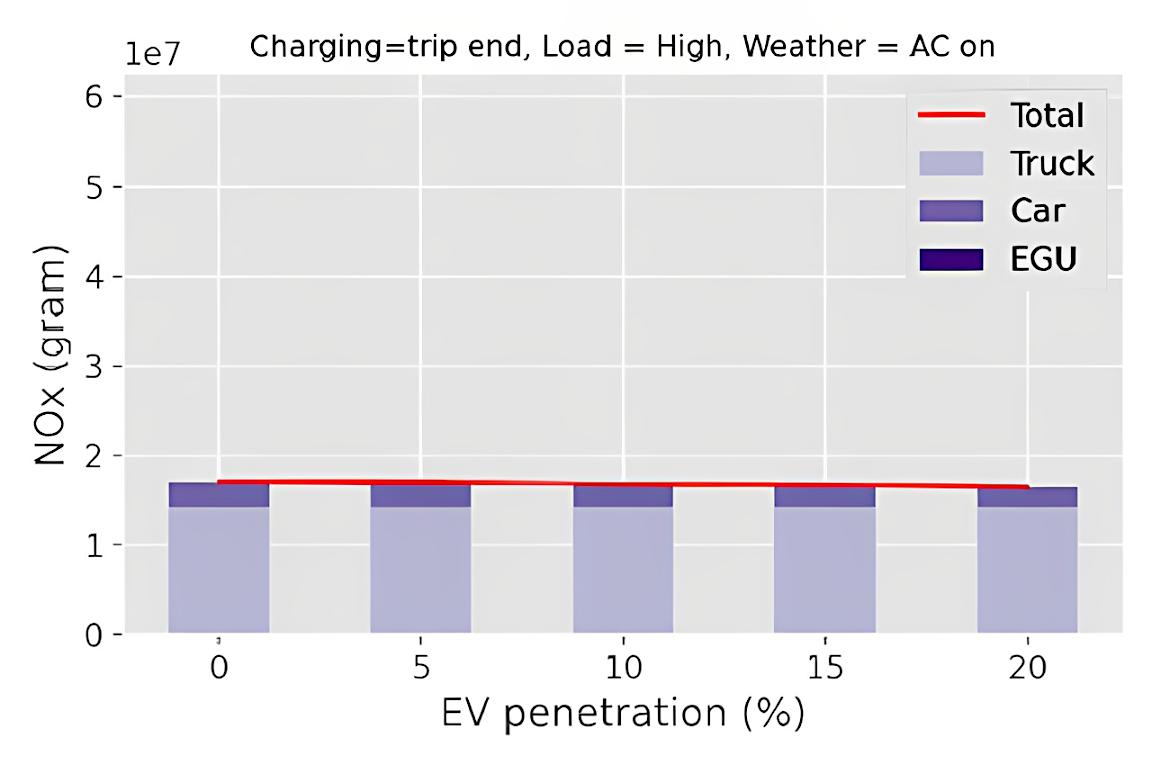}}\label{fig:CO216}
  \hfill
  \subfloat[]{\includegraphics[width=0.45\columnwidth]{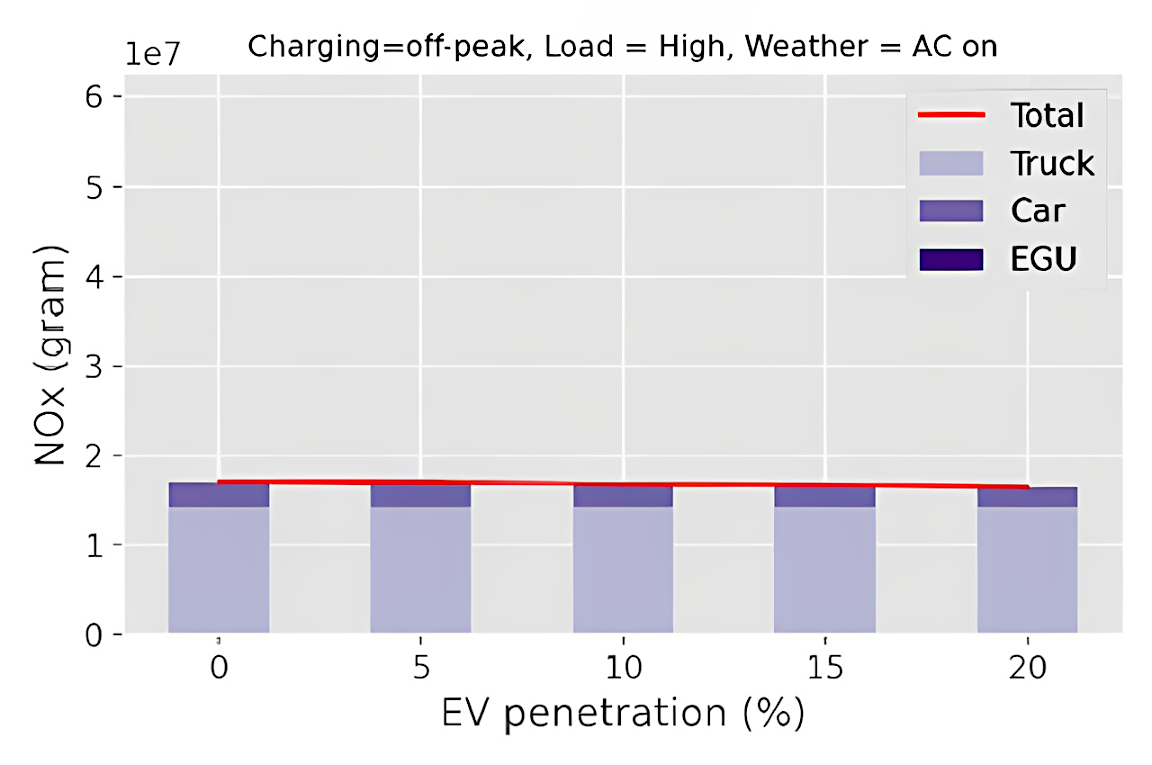}}\label{fig:CO216}
  \hfill
  \subfloat[]{\includegraphics[width=0.45\columnwidth]{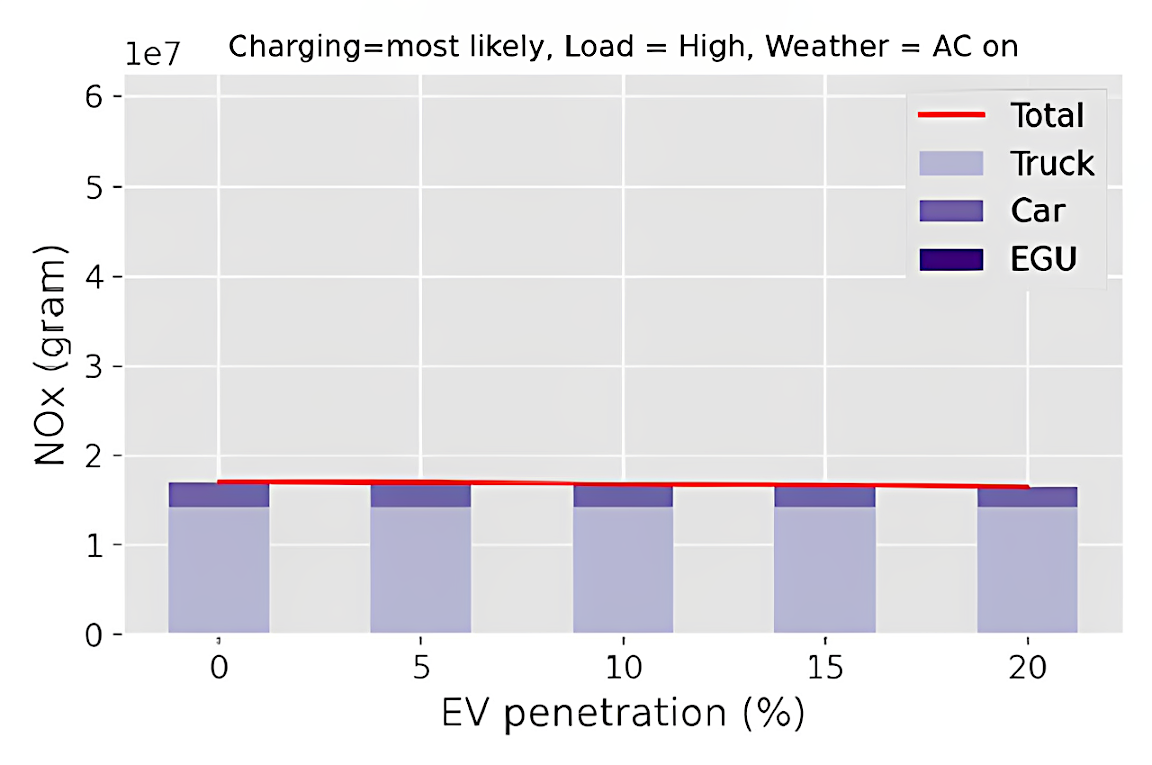}}\label{fig:CO216}
  \hfill
  \caption{Total NOX emissions by scenario (a,b,c) 2016 Case Total NOX Emissions by Scenario (d,e,f) 2030 Case Total NOX Emissions by Scenario}
    \label{fig:NOX}
\end{figure*}

The total CO2 and NOX emissions are used to quantify the environmental impact and the emission results are shown in Figure \ref{fig:CO2} and Figure \ref{fig:NOX} respectively.  Regarding GHG emissions, the total CO2 from on-road and EGUs decreases linearly with the EV market penetration in all cases, with the maximum emission reduction around 15\% under the 20\% EV market share in 2016 case, and 14\% reduction under the 20\% market share in 2030 case.  This finding suggested the elevated emissions from power generation will not offset the benefits of EVs in reducing on-road emissions.  Compared to the on-road emissions, the contribution of EGU emissions is almost negligible in all cases, as low as 0.02\% in 2030 mild weather cases, due to the high efficiency of EVs and power generation.  Most of the GHG difference comes from the on-road emission reduction of light-duty vehicles.
\begin{comment}
   \begin{figure*}[!bp]
  \centering
  \subfloat[]{\includegraphics[width=2\columnwidth]{Figures/2016CO2_high.png}}\label{fig:CO216}
  \hfill
    \subfloat[]{\includegraphics[width=2\columnwidth]{Figures/2030CO2_high.png}}\label{fig:CO230}
  \hfill
  \caption{Total CO2 emissions by scenario (a) 2016 Case Total CO2 Emissions by Scenario (b) 2030 Case Total CO2 Emissions by Scenario}
    \label{fig:CO2}
\end{figure*} 
\end{comment}

Similar to the CO2 trend, the total NOX emissions decreased linearly with the EV market share.  Most of the NOX emissions originate from heavy-duty trucks, with less than 20\% contributed by cars and a negligible amount of emissions by the EGUs.  Under the 20\% EV market penetration and 2016 fuel mix, the total NOX emissions were reduced by only 4\%.  Under the 2030 fuel mix, the total NOX emissions were reduced by 3.5\% under 20\% EV penetration as a result of less NOX contribution of LDVs in 2030 (only 17\% of NOX come from LDVs under no-EV baseline).  Because the fractions of LDV emissions are low, the total NOX emissions in Travis County decreased slightly under various scenarios. 
%The emission reduction benefits will increase if considering start and evaporative emission reductions of EVs, or electrification of HDVs.

\begin{comment}
\begin{figure*}[!bp]
  \centering
  \subfloat[]{\includegraphics[width=2\columnwidth]{Figures/2016NOX_high.png}}\label{fig:NOX16}
  \hfill
    \subfloat[]{\includegraphics[width=2\columnwidth]{Figures/2030NOX_high.png}}\label{fig:NOX30}
  \hfill
  \caption{Total NOX emissions by scenario (a) 2016 Case Total NOX Emissions by Scenario (b) 2030 Case Total NOX Emissions by Scenario}
 \label{fig:NOX}
\end{figure*}    
\end{comment}

%With power generation dominated by natural gas and carbon-free technologies, the EVs are expected to greatly reduce GHG emissions and air pollutants under all proposed scenarios.  The emission reduction benefits achieved are limited to LDVs, while a large portion of emissions come from HDVs.  Therefore, the emission reduction benefits can be improved if a fraction of HDVs can be electrified.  Also, the charging strategies and the baseline electricity load have a substantial impact on the dispatched power generation, and therefore greatly affect the EGU emissions.  
%It is therefore important for system planners and operators in the transportation and the power sectors to coordinate the planning and scheduling of charging events to optimize the system performance, including total costs and emissions.

\section{Conclusions}
In this study, a methodology is developed to asses the environmental impact of EVs under various implementation scenarios.  On the power demand side, a full-chain transportation simulation toolkit called TEMPO is used along with EV charging and energy consumption modules to generate the EV charging load and the on-road emissions under various congestion levels, meteorology, EV penetration levels, and charging strategies.  On the power supply side, the power generation and EGU dispatch are simulated using the OPF method under the combinations of different electric load time series and EV charging loads.  A case study of Travis County, Texas is performed to investigate the total emissions from the on-road and the EGU segments under various scenarios.  The results demonstrate that up to 10\% of the total CO2 can be reduced and up to 4\% of total NOX can be reduced if 20\% of LDVs are EVs.  The EGU emission rates are not sensitive to the EV market penetration and the meteorology scenarios defined in this study, which confirms that higher EV market penetration will likely reduce overall emissions, regardless of other confounding factors.

The emission rate results have revealed complex interactions and trade-offs as the transportation sector electrifies.  Such interactions and trade-offs bear consequential practical implications.  
%For example, the interaction between charging strategy and static electricity load highlights the importance of a carefully designed temporal distribution of charging demand.  
A strategy to nudge charging times off-peak is not always warranted.  Instead, a dynamic pricing signal may be a preferred strategy to modulate charging time depending on the likely EGU dispatch.  

The trade-offs among CO2 and criteria pollutant emissions illuminate the regional policy priorities that decision-makers need to consider in EV deployments.  Because the emission rates of different pollutants can move in opposite directions under the 2016 fuel mix, a strategy that prioritizes VOC reduction may not be the best for a region seeking to reduce NOx.  

Finally, due to the fast decarbonization in the Austin grid from 2016 to 2030, the power generation scenario for each year in this period also changes rapidly.  The emission impacts of EVs in intermediate years need to be investigated using a transitioning power generation mix and may reflect different trends from current results.
%In conclusion, this study has presented an innovative integrated simulation approach unifying transportation and electric power system models.  The integrated simulations have demonstrated the approaches’ advantages in quantifying the environmental impact of transportation electrification under different transportation operations and grid operation scenarios.  The methodology is flexible and modular to model EV’s environmental impact under different non-attainment designations, power generation fuel mix, charging infrastructure availability, and electricity load from other sectors.  Because most MPOs already have their transportation models, and the synthetic grid data is publicly available and ever-expanding \cite{41}, the methodology is readily scalable.  Policymakers and practitioners seeking to answer important ‘what-if’ questions regarding different EV implementation scenarios need such sophisticated tools to find EV deployment solutions that are tailored enough to meet their unique needs yet flexible enough to account for changes in confounding factors. 
\section{Future Work}
The current work can be further expanded and improved in the following ways.  First, due to the lack of real-world EV operation data, the EVs were assumed to follow the operation patterns of conventional vehicles.  Also, the EV charging load is generated under hypothetical charging behaviors.  The real-world EV operation and charging data need to be collected to represent the realistic EV usage patterns.  Second, the emission reduction benefits of EVs may increase if considering start and evaporative emissions.  The modeling approach described in this analysis can be further expanded to include more sources of emissions once activity and emission rates data become available (the regional conformity analysis is currently not needed for Austin attainment area, so those inputs are not available for developing regional emission inventories).  Third, the emissions reduction opportunities in LDV electrification are limited, as shown in our results, so a module to model medium- and heavy-duty vehicle electrification is a logical next step.  Finally, the economic and reliability impacts of EVs on the grid should be evaluated and incorporated into the simulation framework to gain a comprehensive view of EV adoption.  The operation cost and the charging revenue due to EVs, the investment in the EV charging infrastructure, and the potential impact on the grid operations can be added to investigate the cost-effectiveness of EV adoption at the regional level.

\section*{Acknowledgment}
When assisting with this research Jessica Wert and Komal Sheyte were affiliated with Texas A\&M University. Currently, Jessica Wert affiliated with Lawrence Livermore National Laboratory, and Komal Shetye is affiliated Scout Clean Energy.

\Urlmuskip=0mu plus 1mu\relax
\bibliographystyle{IEEEtran}
\bibliography{bibi.bib}

\end{document}